\documentclass[apj]{emulateapj}
\usepackage{color}
\usepackage{epsf}
\usepackage{ulem}
\usepackage{graphicx}

\newcommand{\Om}{\Omega_m}
\newcommand{\Ob}{\Omega_b}

\newcommand{\OL}{\Omega_\Lambda}

\newcommand{\DA}{D\!_A(z)}
\newcommand{\hz}{H(z)}

\newcommand{\hMpc}{h^{-1}{\rm\;Mpc}}
\newcommand{\hGpc}{h^{-1}{\rm\;Gpc}}
\newcommand{\trihGpc}{h^{-3}{\rm\;Gpc^3}}

\newcommand{\ihMpc}{h{\rm\;Mpc^{-1}}}
\newcommand{\kmax}{k_{\rm max}}

\newcommand{\kvec}{\vec{k}}

\newcommand{\al}{\alpha}
\newcommand{\sial}{\sigma_\alpha}

\newcommand{\FoG}{F_{\rm fog}}
\newcommand{\Pm}{P_m}
\newcommand{\nPt}{n_{\rm eff}P_{0.2}}
\newcommand{\sig}{\sigma}

\newcommand{\Sigxy}{\Sigma_{\rm xy}}
\newcommand{\Sigz}{\Sigma_{z}}
\newcommand{\Nb}{{\it N}-body}
\newcommand{\Signl}{\Sigma_{\rm m}}
\newcommand{\Signn}{\Sigma_{\rm nl}}
\newcommand{\Pano}{A(k)}
\newcommand{\Plin}{P_{\rm lin}}
\newcommand{\Plins}{P_{\rm lin,sm}}
\newcommand{\Plinb}{P_{\rm lin,osc}}

\newcommand{\Pnl}{P_{\rm nl}}
\newcommand{\Pnls}{P_{\rm nl,sm}}
\newcommand{\Pnlb}{P_{\rm nl,osc}}

\newcommand{\Psm}{P_{\rm nw}}

\newcommand{\Pmc}{P_{\rm MC}}
\newcommand{\Pmcs}{P_{\rm MC,sm}}
\newcommand{\Pmco}{P_{\rm MC,osc}}

\newcommand{\Cks}{G_{\rm sm}}
\newcommand{\Cko}{G_{\rm osc}}
\newcommand{\Pobs}{P_{\rm obs}}
\newcommand{\Zel}{Zel'dovich}

\newcommand{\don}{d_1}
\newcommand{\dt}{d_2}

\newcommand{\hdelta}{\hat{\delta}}

\begin{document}
\title{High-precision predictions for the acoustic scale in the non-linear
regime}

\author{
Hee-Jong Seo\altaffilmark{1}, Jonathan Eckel\altaffilmark{2}, Daniel J.\ Eisenstein\altaffilmark{2}, Kushal Mehta\altaffilmark{2},  Marc Metchnik\altaffilmark{2}, Nikhil Padmanabhan\altaffilmark{3}, Phillip Pinto\altaffilmark{2}, Ryuichi Takahashi\altaffilmark{4}, Martin White\altaffilmark{5}, Xiaoying Xu\altaffilmark{2}}

\begin{abstract}
We measure shifts of the acoustic scale due to nonlinear growth and redshift distortions to a high precision using a very large volume of high-force-resolution simulations. We compare results from various sets of simulations that differ in their force, volume, and mass resolution. We find a consistency within $1.5-\sigma$ for shift values from different simulations and derive shift $\al(z) -1  = (0.300\pm 0.015)\% [D(z)/D(0)]^{2}$ using our fiducial set. We find a strong correlation  with a non-unity slope between shifts in real space and in redshift space and a weak correlation between the initial redshift and low redshift. Density-field reconstruction not only removes the mean shifts and reduces errors on the mean, but also tightens the correlations. After reconstruction, we recover a slope of near unity for the correlation between the real and redshift space and restore a strong correlation between the initial and the low redshifts. We derive propagators and mode-coupling terms from our \Nb\ simulations and compare with the \Zel\ approximation and the shifts measured from the $\chi^2$ fitting, respectively. We interpret the propagator and the mode-coupling term of a nonlinear density field in the context of an average and a dispersion of its complex Fourier coefficients relative to those of the linear density field; from these two terms, we derive a signal-to-noise ratio of the acoustic peak measurement. We attempt to improve our reconstruction method by implementing 2LPT and iterative operations, but we obtain little improvement. The Fisher matrix estimates of uncertainty in the acoustic scale is tested using $5000\trihGpc$ of cosmological PM simulations from \citet{Taka09}. At an expected sample variance level of 1\%, the agreement between the Fisher matrix estimates based on \citet{SE07} and the \Nb\ results is better than 10 \%.
\end{abstract}

\keywords{
distance scale-- large-scale structure of universe -- methods: numerical
}

\altaffiltext{1}{Center for Particle Astrophysics, Fermi National Accelerator Laboratory, P.O. Box 5
00, Batavia, IL 60510-5011, USA; sheejong@fnal.gov}
\altaffiltext{2}{Steward Observatory, University of Arizona,
                933 N. Cherry Ave., Tucson, AZ 85121, USA}
\altaffiltext{3}{Department of Physics, Yale University, New Haven, CT 06511, USA}
\altaffiltext{4}{Faculty of Science and Technology, Hirosaki University,
  Hirosaki 036-8560, Japan}
\altaffiltext{5}{Departments of Physics and Astronomy, 601 Campbell Hall, University of California Berkeley, California 94720, USA}

\section{Introduction}
In recent years, attention to baryon acoustic oscillations (BAO) as a dark energy probe has increased unprecedentedly due to its robust nature against systematics, and they are now an essential component of most of the major future dark energy surveys under consideration. BAO originate from the sound waves that propagated through the hot plasma of photons and baryons in the very early Universe. At the epoch of recombination, photons and baryons decouple, and as a result, the sound waves freeze out, leaving a distinctive oscillatory feature in the large-scale structure of the cosmic microwave background \citep[e.g.,][]{Mil99,deB00,Han00,Lee01,HalDasi,Netter02,Pearson03,BenoitArcheops,BennettWmap,Hinshaw07,Hinshaw08} and the matter density fields in Fourier space \citep[e.g.,][]{Peebles70,SZ70,Bond84,Holtzman89,HS96,Hu96,EH98,Meiksin99}; In configuration space, the BAO appears as a single spherical peak at its characteristic scale.  The characteristic physical scale of this oscillatory feature, BAO, is the distance that the sound waves have traveled before the epoch of recombination, which is known as ``sound horizon scale''. This sound horizon scale is and will be measured precisely from current and future CMB data. With the knowledge of the physical scale, BAO can be used as a standard ruler to measure angular diameter distance and Hubble parameter at various redshifts and therefore provide critical information to identify dark energy \citep[e.g.,][]{Hu96,Eisen03,Blake03,Linder03,Hu03,SE03}. Recently, BAO have been detected from large-scale structure of galaxy distributions and have been used to place an important constraint on dark energy \citep[][]{Eisen05,Cole05,Hutsi06,Tegmark06,Percival07a,Percival07b,Blake07,Pad07,Okumura08,Estra08,Gazt08a,Gazt08b,Sanchez09,Percival09,Kazin09}. 

Due to nonlinear structure growth at late times, the oscillatory feature of the BAO is increasingly damped, proceeding from small scales to larger scales, with decreasing redshift \citep[e.g.,][]{Meiksin99,SE05,Jeong06,ESW07,Crocce08,Mat08}. Redshift distortions enhance a nonlinear damping along the line of sight direction\citep[e.g.,][]{Meiksin99,SE05,ESW07,Mat08}.  Despite the resulting loss of the BAO signal with decreasing redshift, BAO are believed to be a robust standard ruler. The sound horizon scale is well determined from the CMB and the scale corresponds to $\sim 100\hMpc$ in present time, implying that the feature is still mostly on linear scales where evolutionary and observational effects are much simpler to predict than in the nonlinear regime.  Indeed, such degradation in the contrast of the BAO due to nonlinear structure growth, redshift distortions, and possibly galaxy bias has been studied since the late 90's and is relatively well understood \citep[][]{Meiksin99,Springel05,Angulo05,SE05,White05,Eisen05,Jeong06,Crocce06b,ESW07,Huff07,Smith07,Matarrese07,Nishimichi07,Smith08,Angulo08,Crocce08,Mat08,Taka08,Sanchez08,Jeong09,Taruya09}.  

Nonlinearity also induces a shift of the BAO scale in the low-redshift matter distribution relative to the sound horizon scale measured from the CMB \citep[e.g., ][]{Smith08,Crocce08,Sanchez08}. As the demand for acoustic peak accuracy moves from the current level of a few \% to the sub-percent level of future surveys, we are now required to understand the systematics on the BAO to much better precision. While it is evident that the shift of the BAO scale depends on the choice of the estimator, most recent studies seem to lay more weight on residual shifts using optimal estimators being at a sub-percent level at $z\sim 0$ when accounting for nonlinear structure growth and redshift distortions \citep{Crocce08,Sanchez08,SSEW08}, or even with halo/galaxy bias \citep{Pad09}. Certainly more studies are necessary to confirm the results and ultimately reach a general consensus.

In the previous study \citep{SSEW08} (hereafter SSEW08), we investigated effects of nonlinear evolution on BAO using a large volume (i.e., $320\trihGpc$) of PM simulations. We found that the shift on the acoustic scale indeed increases with decreasing redshift and is less than a percent even at $z=0.3$ in redshift space. 

In this work, we extend the study by using high-force-resolution simulations that are generated by a new \Nb\ code  ABACUS  by Metchnik \& Pinto (in preparation). With the new simulations, we update the evolution of the shifts on the acoustic scale due to nonlinear structure growth and redshift distortions for various force, volume, and mass resolutions. We calculate propagators, i.e., the correlations between the linear density fields and the nonlinear density fields at low redshift, which directly manifest the nonlinear damping of the BAO. We also derive the mode-coupling contribution to the power spectrum with an attempt to qualitatively relate this to the measured values of the shifts. Recently, \citet{PadLag09} showed that the density field after reconstruction is not the linear density field at second order. We relate the propagator and the mode-coupling term to an average and a dispersion of nonlinear density fields in the complex Fourier plane before and after reconstruction, relative to the linear density fields, and derive a signal-to-noise ratio of the standard ruler test from these two terms. The effect of galaxy bias will be presented in companion papers \citep[Mehta et al. in preparation;][]{Xu09}.

It has been demonstrated that the original density-field reconstruction scheme based on the \Zel\ approximation, presented by \citet{ESSS07}, is quite efficient for removing nonlinear degradation on BAO despite its simplicity \citep{SE07,Huff07,SSEW08,PadLag09}. Efficiency of reconstruction in terms of an increase in the signal-to-noise depends on the redshift and the shot noise of the density fields. Meanwhile, it has been shown that the reconstruction removes almost all of the nonlinear shifts of the acoustic scale even when it seemingly is not efficient in terms of the signal-to-noise ratio (SSEW08).  This success, conversely, can be interpreted that a further improvement on the reconstruction scheme will be only a second order effect, at least for BAO. Nevertheless, we discuss  and test possible improvements on our fiducial scheme. While there are other sophisticated reconstruction methods in the literature aimed for the recovery of velocity fields and the initial density fields \cite[e.g.,][]{MAK06},  in this paper we limit ourselves to mild modifications to our fiducial method, mainly due to its proven success for BAO.  We first test an implementation of 2LPT instead of the \Zel\ approximation \citep{Zel70} and second, test iterative operations of the fiducial method in order to improve the signal-to-noise ratio. 

The importance of an accurate prediction of the signal-to-noise ratio for future BAO surveys is evident. The calibration of the Fisher matrix-based estimations against the \Nb\ results have been tried repeatedly, and the resulting discrepancy is at most 20\% (e.g., SSEW08). Further calibration is often limited by the volume of the simulations: in order to minimize {\it the dispersion of the dispersion} among different measurements, we need a large number of random subsamples  while each subsample has an enough cosmic volume to measure the BAO scale.  In this paper, we  calibrate the Fisher matrix estimation based on \citet{SE07} to a level of 1\% by utilizing the enormous cosmic volume of $5000\trihGpc$ from \citet{Taka09}.

This paper is organized as following. In \S~\ref{sec:smethod}, we describe our new cosmological \Nb\ simulations and the methods of $\chi^2$ analysis to measure the acoustic scales from the simulations. In \S~\ref{sec:salphas}, we present the resulting shifts and errors on the measurements of the acoustic scale when accounting for the nonlinear growth and redshift distortions before and after reconstruction. In \S~\ref{sec:spropagators} and \S~\ref{sec:Pmc}, we derive propagators and mode coupling terms from the simulations and qualitatively compare these with the errors and the mean values of the shifts from the simulations. In \S~\ref{sec:StoN}, we relate the propagator and the mode-coupling term to the signal-to-noise ratio of the standard ruler test. In \S~\ref{sec:stLPT}, we implement 2LPT and iteration into our reconstruction scheme and discuss the effect. In \S~\ref{sec:sRyuichi}, we use the $5000\trihGpc$ of simulations by \citet{Taka09} to test the Fisher matrix calculations in \citet{SE07} given the minimal sample variance.
 Finally, in \S~\ref{sec:sdisc}, we summarize the major results presented in this paper.

\section{Methods}\label{sec:smethod}
\subsection{Simulations}

We use three sets of \Nb\ simulations to produce the results presented in this paper. The first two are produced using a new high-force resolution \Nb\ code ABACUS by Metchnik \& Pinto (in preparation) and are used for measuring the shifts of the acoustic scale at different redshifts. Our ABACUS simulations have high force resolution,
sufficient to resolve dark matter halos, unlike the particle-mesh
simulations in SSEW08.  The ABACUS code uses a new method to solve
the far-field gravitational force, resulting in higher force accuracy
than standard tree and Fourier methods, at high computational speed: the code speed is such that each $1024^3$ particle
simulation takes only 3 days on a single 16-core machine.

We generate these two sets using cosmological parameters that are consistent with WMAP5+SN+BAO results \citep{Komatsu09}: $\Om=0.279$, $\OL=0.721$, $h=0.701$, $\Ob=0.0462$, $n_s=0.96$, and $\sig_8=0.817$. The initial conditions are generated  at $z=50$ using the 2LPT based IC code by \citet{Sirko05}: we do not correct for the finite volume effect on the DC mode but assume a zero DC mode.  The two sets differ in their volume and mass resolution. 

The first set employs $576^3$ particles in a box of $2\hGpc$ and a particle mass of $3.2413 \times 10^{12} \,h^{-1}M_\odot$, with a softening length of $0.1736 \hMpc$ (i.e., $1/20$ of the interparticle spacing) for gravity calculation. We generate a total of 63 boxes and therefore a volume of $504 \trihGpc$. We compute power spectra  at $z=3.0$, 2, 1.5, 1, 0.7, and 0.3 using $576^3$ density grids using the cloud-in-cell interpolation. The power spectra are spherically averaged within wavenumber bins of width $\Delta k=0.001\ihMpc$ in real space and redshift space. We will refer to this sample as ``G576''. 

The second set employs $1024^3$ particles in a box of $1\hGpc$ and a particle mass of $7.2110 \times 10^{10} \,h^{-1}M_\odot$, with a softening length of $0.0488 \hMpc$ for gravity calculation. We generate a total of 44 boxes and therefore $44\trihGpc$. The power spectra are computed in the same way as for G576 at $z=1$ (i.e., $\Delta k=0.001\ihMpc$), but using $1024^3$ density grids this time. We will refer to this sample as ``G1024''. We use G1024 to test whether any of the results depend on the different force or mass resolution or simulation volume. 

For the third set, we use the \Nb\ results generated by \citet{Taka09}. This simulation is generated using the Gadget-2 code \citep{SpringelGadget,SpringelGadget2} in a Particle-Mesh (PM) mode with cosmological parameters based on WMAP3 results \citep{Spergel07}: $\Om=0.238$, $\OL=0.762$, $h=0.732$, $\Ob=0.041$, $n_s=0.958$, and $\sig_8=0.76$. The set employs $256^3$ particles in a volume of $1\hGpc$, and therefore a particle mass of $3.94 \times 10^{12} \,h^{-1}M_\odot$, and a $256^3$ force mesh. The initial condition is generated at $z=20$ using \Zel\ approximations, and power spectra are computed at $z=3$, 1, and 0 using $512^3$ grids and are spherically averaged for each discrete value of radial $k$\footnote{The wavenumber is discretized as $k = 2\pi/{L_{\rm box}} \times (n^2_{1}+n^2_{2}+n^2_{3})^{1/2}$ where $L_{\rm box}=1\hGpc$ and $n_i$ (i=1-3) is an integer.}: the width $\Delta k=0$. While this set has worse force resolution, the virtue of this sample is its large volume: $5000\trihGpc$! We will refer to this sample as ``T256''. We utilize this set, first to check the consistency with respect to G576 and G1024, but more importantly to calibrate the Fisher matrix error estimates, that is, to measure to high precision the scatter of the acoustic scale shift. We summarize our three \Nb\ sets in Table \ref{tab:tabsim}.

\begin{figure*}[t]
\epsscale{1.0}
\plotone{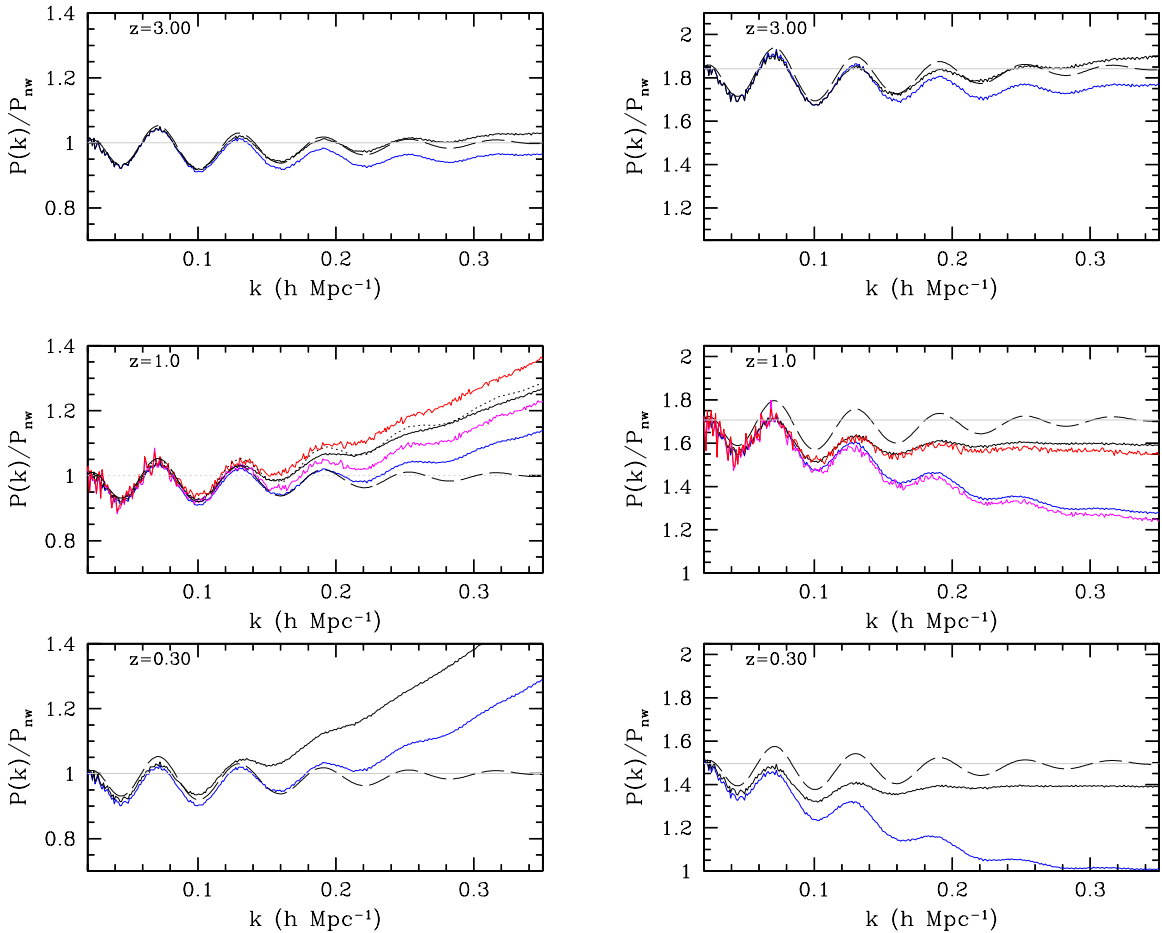}
\caption{Power spectra divided by a smooth, nowiggle power spectrum $\Psm$ at z=3.0, 1.0, and 0.3 in real (left) and redshift space(right), before (black for G576 and red for G1024) and after reconstruction (blue for G576 and magenta for G1024). The dashed line is the linear theory model. The dotted line at $z=1$ is from the halofit model from \citet{Smith03}. We do not subtract shot noise.} 
\label{fig:fP_Psm}
\end{figure*}

\subsection{Fitting methods}
To measure the shift of the acoustic scale, we fit models to the spherically averaged power spectra both in real and redshift space. The fitting method used in this paper is identical to SSEW08. To summarize, we fit the observed power spectra $\Pobs$ to the following fitting formula:

\begin{equation}\label{eq:Pobs}
\Pobs(k)=B(k)\Pm(k/\al)+\Pano,
\end{equation}
where $\al$, $B(k)$, and $\Pano$ are fitting parameters. Here $\al$
is a scale dilation parameter and represents the ratio of the true
(or linear) acoustic scale to the measured scale. For example, $\al>1$
means that the measured BAO being shifted toward larger $k$ relative to
the linear power spectrum. If the nonlinear shift of the acoustic scale were not corrected,
$\al$ would represent the ratio of the mis-measured distance to the true
distance. The template power spectrum $\Pm$ is generated by modifying the BAO 
portion of the linear power spectrum with a nonlinear parameter $\Signl$ to account 
for the degradation of the BAO due to nonlinear growth and redshift distortions as discussed in SSEW08. 
\begin{eqnarray}\label{eq:Pmodel}
\Pm(k)&=&\left[ \Plin(k)-\Psm(k)\right] \exp{\left[ -k^2 \Signl^2/2 \right]} \nonumber \\
&&+\Psm(k),
\end{eqnarray}
where $\Plin$ is the linear power spectrum and $\Psm$ is the nowiggle form from \citet{EH98}. 
We do not include $\Signl$  as a free parameter.
As demonstrated in SSEW08, our results are not sensitive for a wide 
range of $\Signl$ (i.e., $\Delta \Signl = \pm 2 \hMpc$ for $z \lesssim 3.0$) due to the large number of free fitting parameters allowed in $B(k)$ and $A(k)$. We therefore simply choose fiducial values of $\Signl$ based on the \Zel\ approximation from \citet{ESW07}.

The term $B(k)$ allows a scale-dependent nonlinear growth,
and $\Pano$ represents an anomalous power, i.e., additive terms from the
nonlinear growth and shot noise. By including both $B(k)$ and $\Pano$
with a large number of parameters, we minimize the contribution to the
standard ruler method from the broadband shape of the power spectrum. 
We adopt
three different choices of parametrization for $B(k)$ and $\Pano$. 
For real space, we use the following two choices.
For the first choice, which we call ``Poly7'', we use a quadratic polynomial in $k$ for $B(k)$ and 
a $7^{th}$ order polynomial function for $A(k)$. For the second choice, referred to as ``Pade'',
we use Pade approximants for $B(k)$ in a form of $b_0(1 + c_1 k + c_3 k^2 + c_5 k^3)/(1 + c_2 k + c_4 k^2)$, 
and a quadratic polynomial in $k$ for $A(k)$.

For the redshift space fits, we extend the parametrization in equation (\ref{eq:Pobs}) to include a finger-of-God (FoG) term.
\begin{eqnarray}\label{eq:redout}
P(k)&=&[ B(k)P_m(k/\al) + \Pano] \times \FoG\\
&+& e_1,\nonumber\\
\mbox{with }
\FoG&=&\exp\left[-(k \don)^{\dt}\right]. \nonumber\\
\end{eqnarray}
Here $B(k)$ and $A(k)$ are quadratic polynomials in $k$, and $\don > 0\hMpc$ and $\dt>0$.
We call this ``Poly-Exp-Out''. We use a fitting range of $0.02\ihMpc <k<0.4\ihMpc$. Our results are highly consistent for $ 0.35\ihMpc <\kmax < 0.5\ihMpc$ and $k_{\rm min} < 0.04 \ihMpc$ where $k_{\rm min}$ and $\kmax$ are the two outer bounds of the fitting range.

As demonstrated in SSEW08, the fits are virtually the same for different choices of fitting formulae given the sample variance. While we try variations in $\Signl$ and fitting formulae for all the power spectra we generated, 
we only quote values derived using Poly7 in real space and using Poly-Exp-Out in redshift space.

\subsection{Resampling methods}
The true covariance matrix of the \Nb\ simulations is unknown a priori. Therefore we use the variation between the simulations to assess the true scatter
in $\al$. We fit each simulation assuming that the covariance matrix for $P(k)$ 
is that
of a Gaussian random field, i.e., assuming independent band powers with
variances determined by the number of independent modes in each. Although
this is not an optimal weighting of the data for the determination of
$\al$, the effects from non-Gaussianity in the density field will still 
be reflected in the scatter between best-fit $\al$'s from different simulations \citep[also see][for the negligible effect of non-Gaussian errors in multi-parameter fitting]{Taka10}.

We use various resampling methods to measure the shifts and the scatters of shifts of the acoustic scale.
In the first method, we generate 1000 subsamples by randomly resampling M boxes out of total N simulations without replacement. We perform a $\chi^2$ analysis to the individual subsamples and find the mean and the scatter in the best fit $\alpha$. The scatter in $\alpha$ is rescaled by $\sqrt{M}/\sqrt{N-M}$ to give a scatter associated with the mean value of $\alpha$ for a total of N simulations, and is presented in Table \ref{tab:recreal}. We choose $M=31$ for G576 ($N=63$) and $M=22$ for G1024 ($N=44$). Note that $M \simeq N/2$, and now the scatter of the subsamples (presented in Figure \ref{fig:fal_al576_0.3}, \ref{fig:fal_al576_1.0}, and \ref{fig:fal_al1024_1.0}) is very similar to the scatter associated with the mean value of $\alpha$ (in Table \ref{tab:recreal}).

We also use jackknife resampling to measure the shifts and find that both methods give consistent estimates of the shift and the scatter of the shift for G576 and G1024.  For T256, we measure the shifts primarily by using the jackknife resampling; we also use a third resampling method, Bootstrap resampling, to measure the scatters among 125 subsamples of 40 boxes and 250 subsamples of 20 boxes. For T256, the means of the shifts from the two resampling methods are almost identical and the errors on the shifts agree within their expected uncertainties.

\begin{figure*}
\plotone{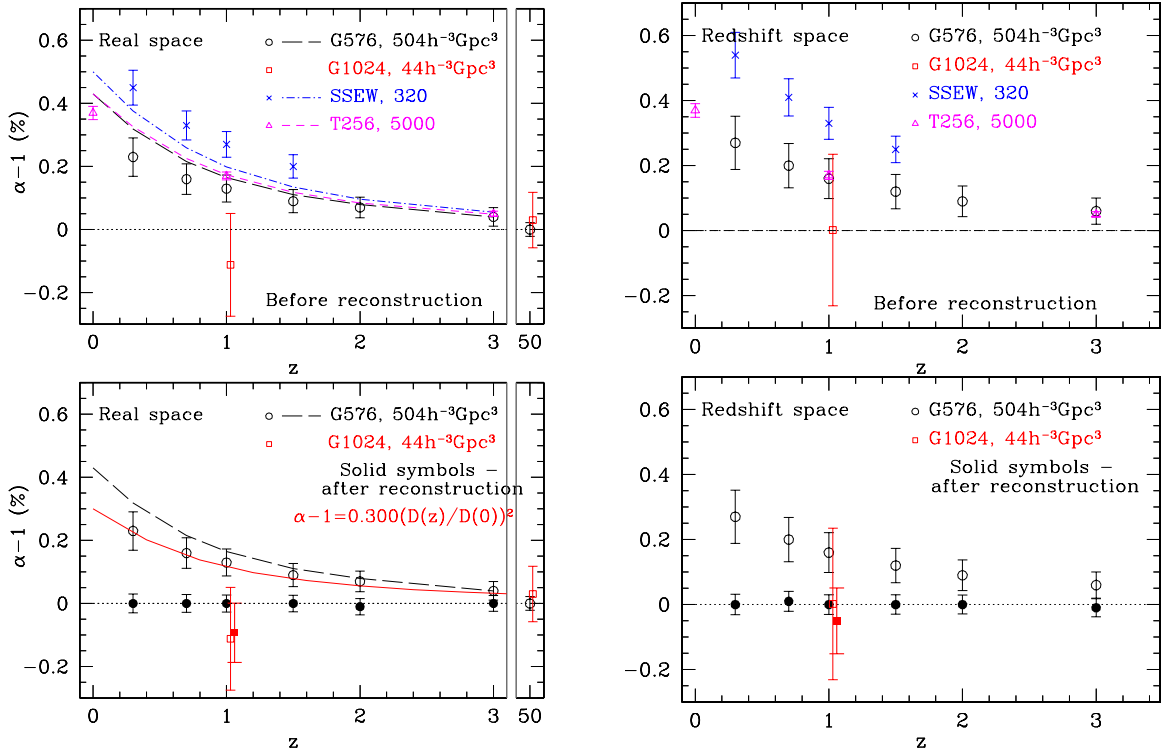}
\caption{The nonlinear evolution of shifts $\al-1$ with redshift. Top panels show the mean and the errors of the mean of $\al$ before reconstruction. The broken lines with the corresponding colors are the expected nonlinear shifts derived using the second order perturbation theory, as in \citet{Pad09}. In the bottom panels, solid points show the values after reconstruction, in comparison to the values before reconstruction (open points).  Data points for G1024 are slightly displaced in $z$ for clarification. We note that the sample variance is highly correlated between shifts at different redshifts for a given set of simulations. The solid red line in the bottom panel is our best fit to $\al(z)-1 \propto [D(z)/D(0)]^2$ when including the covariance between redshifts.}
\label{fig:falphas}
\end{figure*}

\subsection{Reconstruction method}
We reduce degradations in the BAO due to nonlinear growth and redshift distortions using the density-field reconstruction method described in \citet{ESSS07}. We smooth gravity of the observed nonlinear density fields on small scales with a Gaussian filter, apply a linear perturbation theory continuity equation $\bigtriangledown \cdot \vec{q} = - \delta $, and derive the linear theory motions $\vec{q}$. We displace the real particles and the uniformly distributed reference particles by $-\vec{q}$, derive the two density fields, and find the final, reconstructed density field by subtracting the two. We use $R=10\hMpc$ as our fiducial width of the Gaussian filter. In redshift space, the displacement field derived from the observed redshift-space density field is multiplied by $(1+f)/(1+\beta)$ along the line of sight in order to approximately account for the linear bias and the redshift distortions. For mass, $(1+f)/(1+\beta)=1$, meaning that the redshift-space density field exactly accounts for the redshift distortions in linear theory \citep{Kaiser87}\footnote{Here $f$ is the logarithmic derivatives of the linear growth factor and $\beta=f/bias$.}. We do not correct for the FoG compression, but the effect of the FoG compression is briefly discussed in \S~\ref{sec:StoN}.

\section{The evolution of acoustic scales}\label{sec:salphas}
We show the spherically averaged real-space and redshift-space power spectra of the matter distribution for G576 and G1024 in Figure \ref{fig:fP_Psm}.  As expected from its higher mass resolution, G1024 (at $z=1$) shows  slightly higher power on small scales than G576 in real space while showing more FoG effect in redshift space.

\begin{figure*}
\plotone{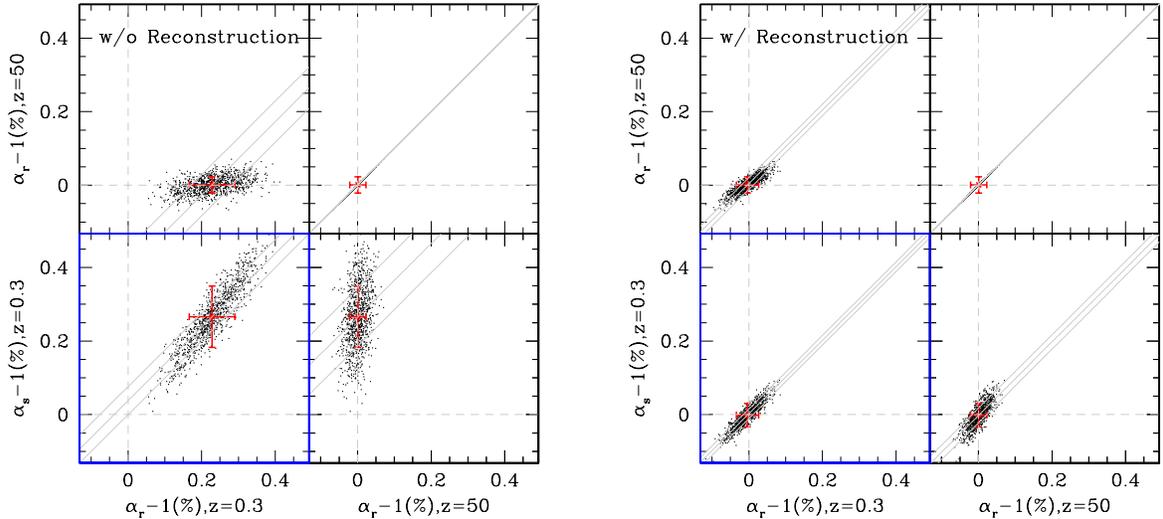}
\caption{$\al-1$ of 1000 subsamples from 63 boxes of G576 at $z=0.3$,  before reconstruction (left) and after reconstruction (right). Each subsample is generated by a random resampling of $M=31$ boxes out of the total $N=63$ boxes without replacement. $\al_r-1$ denotes shift values in real space and $\al_s-1$ is for redshift space. The red error bars denote the mean and the standard deviation of $\al-1$. Note that the standard deviation among subsamples closely represents the scatter associated with the mean of $\alpha$ as $M \sim N/2$.  The gray diagonal lines are graphical representation of the distribution of the differences in $\alpha$'s: the y-intercept or x-intercept of the middle gray line shows $\Delta \al$ and the outer gray lines depict a $1-\sig$ range for $\Delta \al$.  }
\label{fig:fal_al576_0.3}
\end{figure*}

\begin{figure*}
\plotone{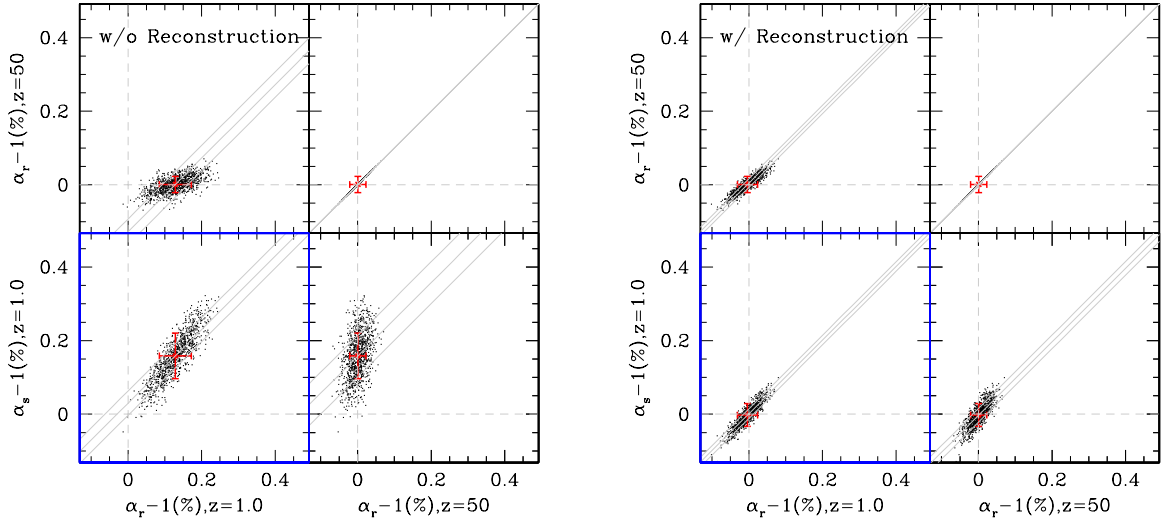}
\caption{ $\al-1$ for all the subsamples for G576 at $z=1.0$. The red error bars denote the mean and errors of $\al-1$.}
\label{fig:fal_al576_1.0}
\end{figure*}

\begin{figure*}
\plotone{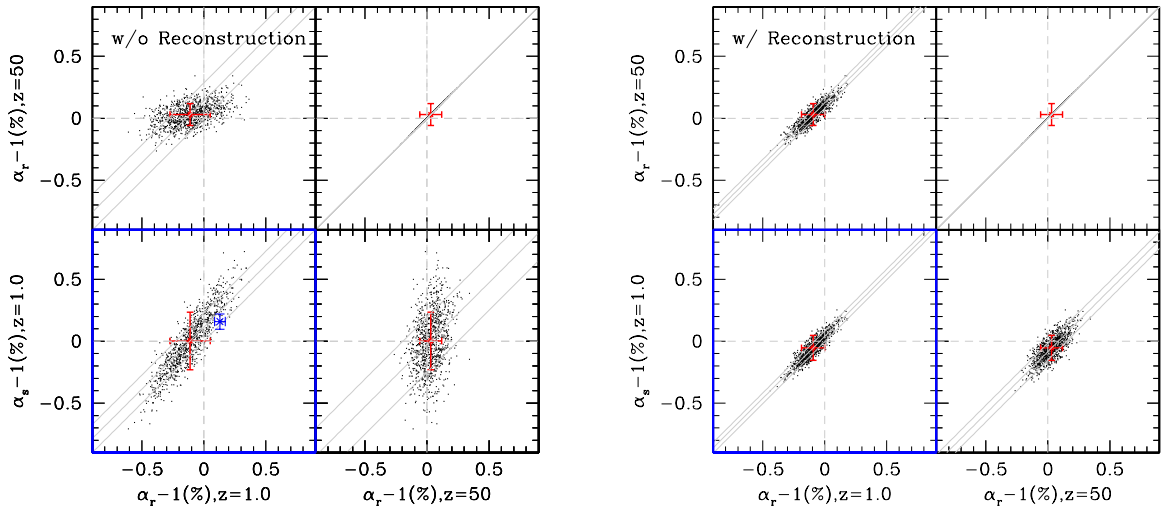}
\caption{$\al-1$ for all the subsamples for G1024 at $z=1.0$ (black points) before reconstruction (left) and after reconstruction (right). The red error bars denote the mean and errors of $\al-1$.  The standard deviations of the subsamples is equal to the scatter associated with the mean of $\alpha$. We superimpose the result of G576 at  $z=1$ with a blue error bar, for comparison. }
\label{fig:fal_al1024_1.0}
\end{figure*}

\begin{figure}[b]
\plotone{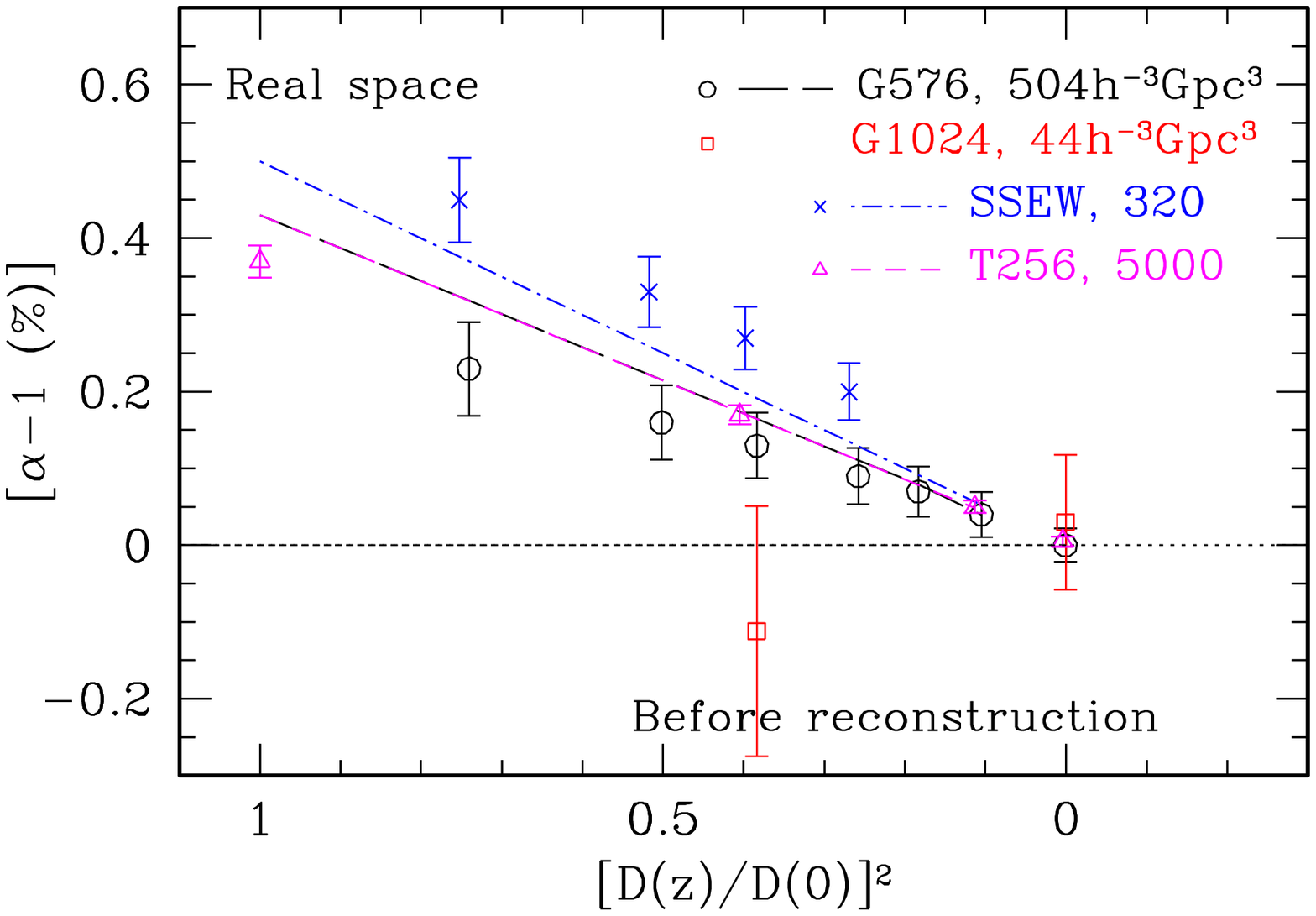}
\caption{We replot the top left panel of Figure \ref{fig:falphas} as a function of $D^2(z)$, where $D(z)$ is the linear growth factor, in order to absorb differences due to different growth factors between cosmologies. The 2PT predictions (broken lines) appear as straight lines in this case. The different slopes are due to the different nonlinear structure growth (i.e., effects of the second order term on the BAO) depending on cosmology. The predictions for T256 (short-dashed magenta) and G576 (long-dashed black) overlap. Note that a fair fraction of the difference between SSEW and the other sets is explained by the difference in cosmologies. }
\label{fig:falDz}
\end{figure}

Figure \ref{fig:falphas} and Table \ref{tab:recreal} show the resulting mean and the errors of the mean of the shifts from the subsamples, in comparison to the results from T256 and previous results from SSEW08. We find that the shifts measured from G576 are less than those measured in SSEW08 by up to $\sim 2.7-\sig$. However, the cosmologies in the two sets are not identical, which we need to take into account. We derive a cosmological scaling prediction for shifts using the second order perturbation theory (hereafter 2PT) based on \citet{Pad09} and compare the predictions (broken lines with the corresponding color in the figures) with the measurements. Indeed the shifts predicted for the cosmology of SSEW08 are higher than for G576. Then the disagreement between SSEW08 and the prediction is about $1.8-\sigma$.

The measured shifts for G576 and T256 are in a good agreement (i.e., within $1-\sigma$ at the shared redshift bins) and the 2PT prediction matches the numerical result to a reasonable level, especially for $z\gtrsim 1$. The agreement appears worse at low redshift, which might be relevant to 2PT being a better approximation at high redshift. Meanwhile, the shift at $z=1$ from G1024 is $1.5-\sigma$ smaller than G576. In redshift space, the shift values from G1024 are bigger than that in real space, as expected, and are in a better agreement with other sets than in real space (i.e., within $1-\sigma$). As the sample variance of G1024 is relatively large compared to the other sets, we suspect that the discrepancy is largely due to the statistical fluctuations rather than a resolution or simulation volume.

The figure illustrates that our reconstruction scheme based on \citet{ESSS07} reduces the shifts essentially to zero in most of the cases, while the shift is consistent with zero at a $ 1-\sig$ level for G1024 in real space.  The errors on shifts have decreased by a factor of 2 (2.6) at $z=0.3$, a factor of 1.5 (2) at $z=1$, and a factor of 1.2 (1.4) at $z=3$ in real space (in redshift space).

Left panels of Figure \ref{fig:fal_al576_0.3} and \ref{fig:fal_al576_1.0} show the distributions of  $\al-1$ for all the subsamples for G576 before reconstruction. An intriguing feature is the strong correlation between $\al-1$ in real space and $\al-1$ in redshift space in the lower left corners of the figures. The slope of the correlation is larger than unity and this implies that shifts in redshift space are greater than shifts in real space. The strong correlation and the non-unity slope is expected, as the displacements in redshift space are correlated with but larger than those in real space \citep{ESW07}. Meanwhile, there seems no obvious correlation of $\al-1$ between the low redshift and the initial redshift. While not shown in the figure, the correlation between any two redshifts becomes stronger for a smaller redshift difference.

With reconstruction, the right panels of the figures show that we recover a slope of near unity for the correlation between the real and redshift space $\al-1$. Also note that we recover a strong correlation between the low and the initial redshifts. The nonlinear growth involves a second-order process that is imperfectly correlated with the initial density fields, and the former dominates over the shifts imprinted by the initial condition  \citep{Crocce08,Pad09}. As the reconstruction removes shifts due to this second-order process, i.e., due to nonlinear structure growth, the remaining shifts are dominated by the variations in the initial conditions. 

Because the various $\alpha$'s within individual subsamples are highly corrected, it is effective to study the differences of $\alpha$'s, as the errors on the differences is considerably smaller than the quadrature sum of the two terms. This allows us to study certain trends with reduced sample variance. We derive the differences of $\al$, that is, $\Delta \al_{sr}=\al_{s}-\al_{r}$, where $\al_s$ is for redshift space and $\al_r$ for real space, and $\Delta \al_{z,z50}=\al_{z}-\al_{\rm z50}$, where $\al_z=\al_r$ or $\al_s$, for each subsample and calculate the mean and the errors on the mean. The results are listed in Table \ref{tab:recreal} and are also graphically represented with the gray diagonal lines in Figure  \ref{fig:fal_al576_0.3} and \ref{fig:fal_al576_1.0}: the y-intercept or x-intercept of the middle gray line shows $\Delta \al$ and the outer gray lines depict a $1-\sig$ range for $\Delta \al$. The results in Table \ref{tab:recreal} are consistent with the trends of correlation we observed in the figures: errors on $\Delta \al_{sr}$ are indeed less than a quadratic sum of $\al_s$ and $\al_r$, implying a strong correlation between shifts in real and redshift space. Errors on $\Delta \al_{z,z50}$ before reconstruction, on the other hand, are very similar to a quadratic sum, which is consistent with the little correlation observed in the figures. After reconstruction, errors on all differences become much smaller than the quadratic sums.

Figure \ref{fig:fal_al1024_1.0} shows results for G1024 (black points and red error bars) that are very similar to G576 (blue error bar) except that the mean of the distribution seems biased relative to G576 by $1.5-\sig$. We believe that sample variance can be a reasonable cause for $1.5\sig$ discrepancy between G576 and G1024. Obviously, reconstruction cannot correct shifts in this case: note that the sample variance on shifts after reconstruction is at the level of the initial condition, and reconstruction cannot do anything to the sample variance in the initial conditions.

Finally, we attempt to model the growth of $\al$ with redshift as a function of the linear growth factor $D(z)$. In order to reduce the sample variance effect from the initial condition, we use $\Delta \al_{z,z50}$ to derive the growth of $\al$. From the results of G576, when we perform a $\chi^2$ analysis to the mean $\al$ at $z=0.3$, 1, 3.0 with the covariance derived from the resampling, we find that $\al(z) -1 = (0.295\pm 0.075)\% [D(z)/D(0)]^{1.74\pm 0.35}$.
The resulting power index is in agreement with the expected $D^2(z)$ dependence from the 2PT within $1-\sigma$, and is in excellent agreement with what \citet{Pad09} measured using their $\chi^2$ fitting, despite the different cosmology and different simulation sets used. It is also in good agreement with the power index measured from SSEW08. If we fix the power index to be $2$, as expected from the 2PT, we find the best fit of $\al(z) -1  = (0.300\pm 0.015)\% [D(z)/D(0)]^{2}$ (solid line in Figure \ref{fig:falphas}).

Figure \ref{fig:falDz} shows $\al$ in real space as a function of $D^2(z)$ by which we absorb differences due to different growth factors between cosmologies. The 2PT predictions appear in straight lines in this case and the different slopes are due to the different nonlinear structure growth (i.e., effects of the second order term on the BAO) depending on cosmology. Now the predictions for T256 and G576 are almost the same. Again, G576 and T256 follow a $D^2(z)$ dependence at high redshifts (i.e., at small values of $D(z)^2$) while the dependence becomes weaker at low redshifts.

In redshift space, the displacements of mass tracers are larger along the line of sight direction by $(1+f)$ \citep{ESW07}, and we therefore expect more shift along the line of sight by the amount that depends on $f$, where $f$ is the logarithmic derivatives of the linear growth factor ($\sim \Om(z)^{0.6}$). From G576, we derive shifts in redshift space that are greater than real space by 34\%, 23\%, and 17\% at $z=3$, 1, and 0.3, respectively. This roughly agrees with the shift increase of $\sim 1+f/3 = 33\%$, 30\%, 25\% at $z=3$, 1, and 0.3 that would be expected for a spherically averaged redshift space power spectrum if the shift along the line of sight were $(1+f)$ times larger.


\section{Evolution of Propagator}\label{sec:spropagators}

\begin{figure*}
\plotone{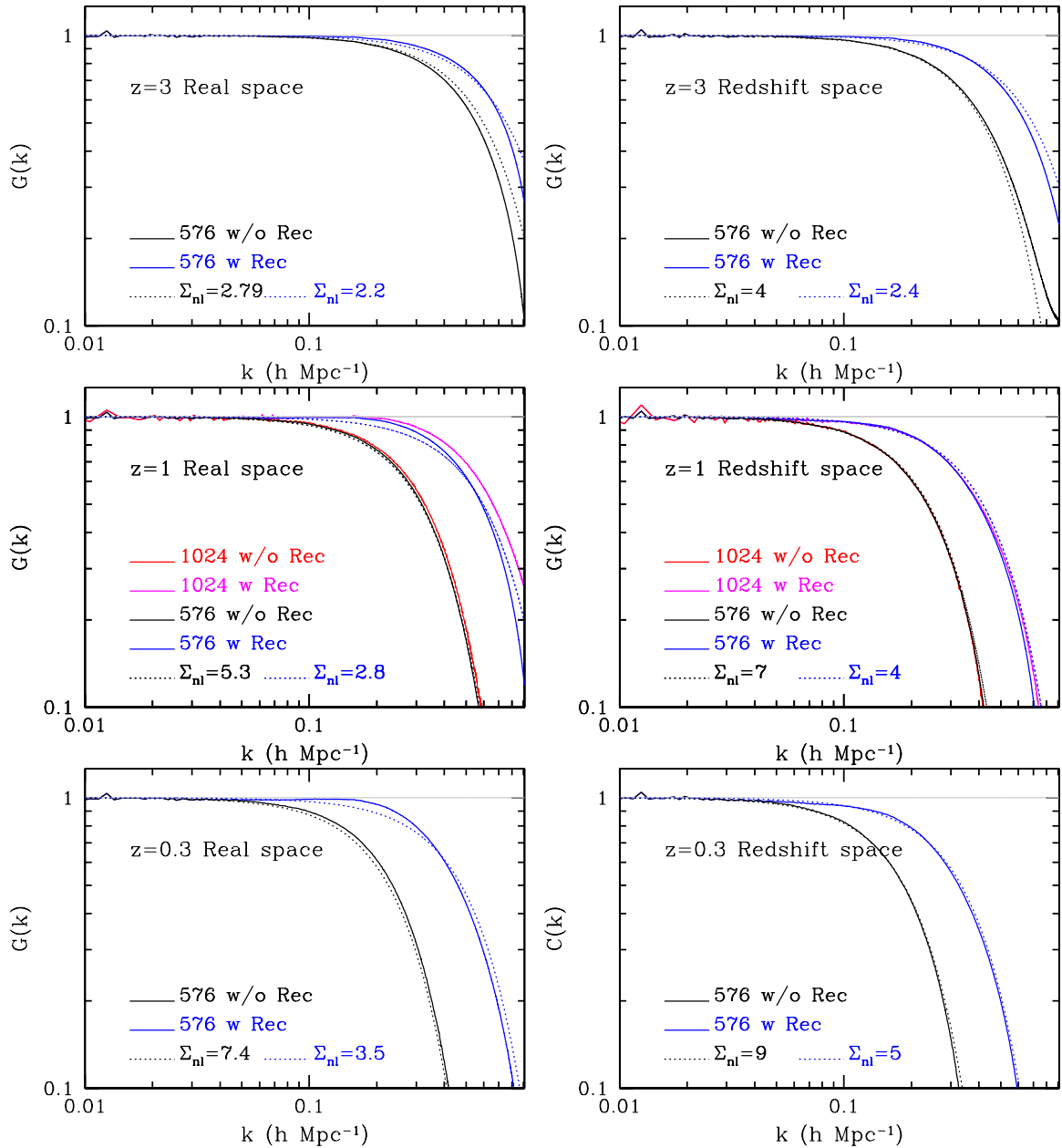}
\caption{$G(k)$ at z=3.0, 1.0, and 0.3 in real (left) and redshift space(right), before (black for G576 and red for G1024) and after reconstruction (blue for G576 and magenta for G1024). The Gaussian model of $G(k)$ is over-plotted with dotted lines: $\Signn$ for real space before reconstruction is chosen based on the \Zel\ approximation in \citet{ESSS07}. }
\label{fig:fCk}
\end{figure*}

The degradation of BAO in the final density fields, before reconstruction, can be reasonably approximated by a linear correlation function convolved with a Gaussian function in configuration space or equivalently a Gaussian damping of the linear BAO peaks in Fourier space \citep{Eisen05,Crocce06b,ESW07,Crocce08,Mat08,PadLag09}, based on the \Zel\ approximation. Meanwhile, it has not been investigated whether such an approximation is a good description for the density fields after reconstruction \citep[but see][]{PadLag09,Noh09}.  We can assess the exact amount of the remaining BAO signal in the final density fields in a numerical way by deriving  the cross-correlation between the initial and final density fields, i.e., ``propagator'' \citep{Crocce06b,Crocce08}. In this section, we derive propagators (eq. [\ref{eq:propa}]) for real and redshift space and before and after reconstruction: 

\begin{equation}\label{eq:propa}
G(k,z)=\frac{<\hdelta_{\rm lin}(\kvec,z)\hdelta(\kvec',z)>}{<\hdelta_{\rm lin}(\kvec,z)\hdelta_{\rm lin}(\kvec',z)>},
\end{equation}
where $\hdelta_{\rm lin}(\kvec,z)$ is the initial linear fields that is linearly scaled to $z$, and $\hdelta(\kvec,z)$ is from the final density fields at $z$.  
This implies that the nonlinear power spectrum can be modeled as \citep{Crocce08}
\begin{equation}\label{eq:Pnl}
\Pnl(k,z)=G(k,z)^2 \Plin (k,z) + \Pmc (k,z),
\end{equation}
where $\Plin$ is the input power spectrum that is linearly scaled to $z$, and $\Pmc$ is the mode-coupling term that describes a portion of power spectrum that is not directly correlated with the initial fields. 

Figure \ref{fig:fCk} shows $G(k)$ in various cases. The black solid lines and the dotted lines, respectively, show the propagators from our simulations and the Gaussian damping model based on the analytic \Zel\ approximation from \citet{ESW07}\footnote{We derive the estimates of nonlinear damping parameter $\Signn$ for the pairs separated by the sound horizon scale.}. In redshift space (right panels), unlike in real space (left panels), we use a by-eye estimate of an isotropic $\Signn$ to represent an anisotropic damping model. Before reconstruction, based on the real-space $G(k)$, the \Zel\ approximation is a good  description of  $G(k)$ for the region of $0.1<G(k)<1$, except for $z=3$ . Meanwhile, it is difficult to estimate the amount of the remaining Gaussian damping after reconstruction, as the effectiveness of the reconstruction will vary depending on redshifts as well as shot noise: for example, at high redshift, there is less nonlinearity from which to recover BAO. The blue solid lines in the figure show the propagators from our simulations after reconstruction, and blue dotted lines show our crude by-eye estimation of the corresponding Gaussian damping model. 
Although the Gaussian damping model appears to be a worse description for $G(k)$ after reconstruction, we estimate that the decrease in the nonlinear damping parameter due to reconstruction is roughly a factor of 2 at $z=0.3$. At $z=1$, we show both G576 and G1024. Note that the propagators are very similar before reconstruction despite the different force, mass, and volume resolution: it differs only by $3 \sim 4\%$ at $k=0.2\ihMpc$. The difference in the propagators is bigger after reconstruction: the reconstruction seems somewhat more effective in G1024 in real space, probably due to its slightly higher amplitude on quasilinear scale and smaller shot noise (Figure \ref{fig:fP_Psm}).


\section{The mode-coupling term, $\Pmc$} \label{sec:Pmc}
 
\begin{figure*}
\plotone{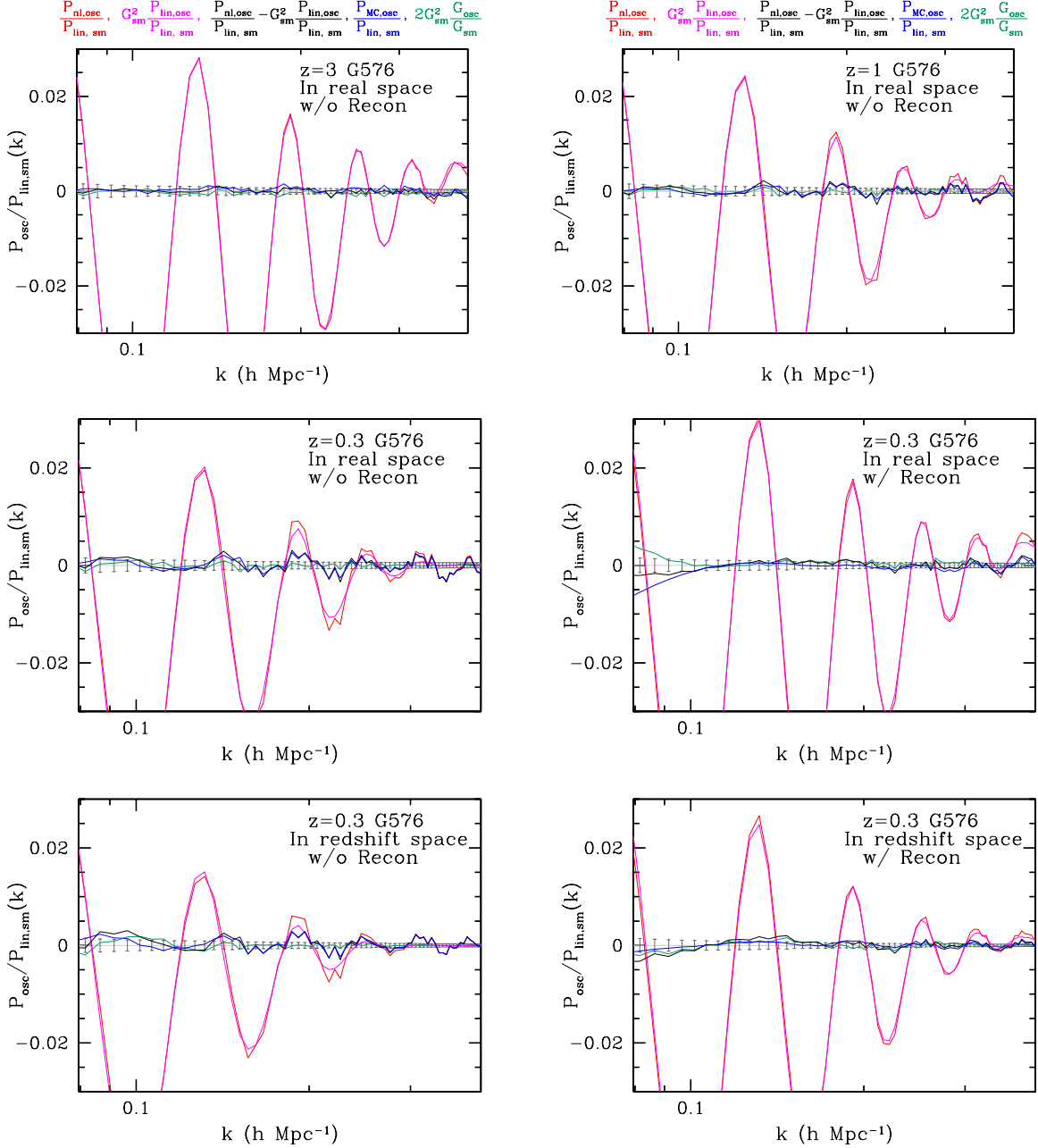}
\caption{The source of nonlinear shift: we show non-smooth components of $\Pnl$ for G576.  The first three panels shows G576 at various redshifts before reconstruction, and the middle-right panel shows G576 at $z=0.3$ after reconstruction. The bottom panels show G576 in redshift space before and after reconstruction. The red line shows the observed nonlinear BAO, $\Pnlb/\Plins$, and the magenta shows the degraded BAO without a nonlinear shift, $\Cks^2\Plinb/\Plins$. Note that, without reconstruction, the amplitude of BAO in the red and magenta lines decrease with decreasing redshift. The black is the difference between the red and the magenta, $\Pnlb/\Plins-\Cks^2\Plinb/\Plins$, therefore the source of the observed nonlinear shift. Note that the oscillatory, non-smooth feature in the black lines indeed increases with decreasing redshift. The blue line is for the non-smooth contribution from the mode-coupling term, i.e. $\Pmco/\Plins$, and the green is the non-smooth contribution from the propagator, $2\Cks^2(\Cko/ \Cks)$. The features in the blue lines appear in good agreement with the features in the black lines and therefore indeed seem responsible for the observed shift. The gray errors around zero show statistical fluctuations expected in $\Pnl$. We used $\Delta k=0.005 \ihMpc$ to decrease a sample variance. The broad-band deviation of components from zero on small wavenumbers in the reconstructed cases is due to a poor fitting to the smooth components. }
\label{fig:fPmc}
\end{figure*}

In the previous section, we investigated the propagator, which is a description of the signal in the signal to noise ratio associated with the acoustic scale measurement. In this section, we investigate the source of the observed shift of the acoustic scale shown in Figure \ref{fig:falphas}.
Assuming that the propagator is smooth, the source of the observed shift should reside in $\Pnl(k,z)-G(k,z)^2 \Plin (k,z)$, and therefore the mode-coupling term $\Pmc$, by its definition in equation (\ref{eq:Pnl}). Here $\Pnl(k,z)$ is the measured nonlinear power spectrum with its acoustic scale shifted relative to the linear power spectrum, and $G(k,z)^2\Plin$ is a nonlinear power spectrum with the degraded BAO but without the shift. Therefore the difference between these two power spectra will show the source of the shift. More specifically, we are interested in oscillatory features in $\Pnl(k,z)-G(k,z)^2 \Plin (k,z)$ and $\Pmc$ that are out of phase in wavenumber relative to the BAO and therefore mimic the derivative of the BAO with respect to wavenumber.

\begin{figure*}
\plotone{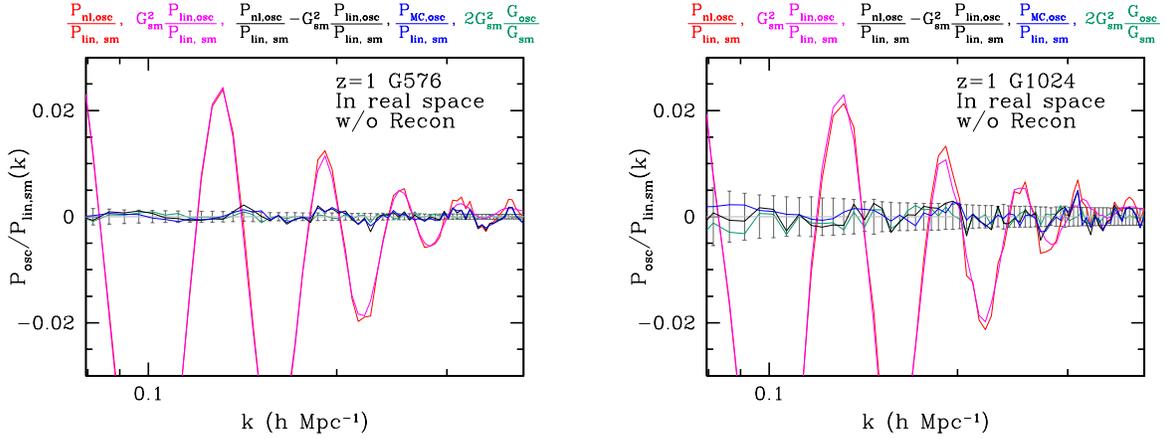}
\caption{Non-smooth components of $\Pnl$ at $z=1$ for G575 and G1024. The left panels show G576 and the right panel shows G1024. The different small-scale clustering shown in Figure \ref{fig:fP_Psm} suggests that the overall mode-coupling term is slightly different between G1024 and G576. We however do not find an indication for a difference in the {\it non-smooth} component of $\Pmc$ between G1024 and G576 induced by a difference in force and mass resolution or size of the box used for the simulations.}
\label{fig:fPmc1024}
\end{figure*}

As we are interested in the oscillatory components only, we rewrite equation (\ref{eq:Pnl}) and separate the smooth and non-smooth components in $\Pnl$, $G(k)$ (if any), and $\Pmc$.
Then
\begin{eqnarray}
\Pnl&=&\Pnls \left(1+\frac{\Pnlb}{\Pnls} \right)  \\
&=& \Cks^2 \left(1+\frac{\Cko}{\Cks}\right)^2 \times \Plins\left(1+\frac{\Plinb}{\Plins}\right)\nonumber\\
&+&\Pmcs \left( 1+\frac{\Pmco}{\Pmcs} \right),
\end{eqnarray}
where $\Pnls$, $\Plins$, $\Cks$, and $\Pmcs$ are the smooth components and $\Pnlb$, $\Plinb$, $\Cko$, and $\Pmco$ are any non-smooth components that include oscillatory components as well as random noise. Note that we have not assumed that $\Cks$ is smooth.
We remove all additive terms that involve only smooth components, as these can be marginalized over, and then keep only the first order term in $\Cko/\Cks$.  Finally, we divide all terms with $\Plins$:
\begin{eqnarray}
\frac{\Pnlb}{\Plins}&\sim& \Cks^2\frac{\Plinb}{\Plins}+2\Cks^2\frac{\Cko}{\Cks} + \frac{\Pmco}{\Plins}.\label{eq:Pmc}
\end{eqnarray}
Therefore the nonlinear shift that is imprinted in $\frac{\Pnlb}{\Plins}- \Cks^2\frac{\Plinb}{\Plins}$ will depend on $\frac{\Pmco}{\Plins}$, as long as $\Cko\sim 0$. Our goal is first to show the source of the observed nonlinear shift in $\frac{\Pnlb}{\Plins}- \Cks^2\frac{\Plinb}{\Plins}$ and second, to compare it with $\frac{\Pmco}{\Plins}$ after inspecting $2\Cks^2\frac{\Cko}{\Cks}$ for any obvious oscillatory feature.

In detail, we derive the smooth components, $\Pnls$, $\Plins$, $\Cks$, and $\Pmcs$ in the following ways. For $\Pnls$, we use equation (\ref{eq:redout}) but with $P_m$=$\Psm$ where $\Psm$ is the nowiggle form from \citet{EH98}. We then recycle the best fit $\alpha$, $B(k)$, and $A(k)$ derived for $\Pnl$ to produce $\Pnls$ because directly fitting to $\Pnls$, which does not have any BAO feature, produces degeneracies among the fitting parameters. We follow the same procedure for $\Plins$ using the randomized power spectrum at $z=50$. In order to generate $\Cks$, we combine a Gaussian damping model with a Pade approximant and polynomials\footnote{$\Cks(k)=d_0 \exp\left[-d_1 k^2\right]\frac{(1+c_1 k + c_2 k^2)}{(1+b_1 k)}+(a_0+a_1 k +a_2 k^2)$ where $a_i's$, $b_i's$, $c_i's$, and $d_i's$ are fitting parameters.}. Finally we derive $\Pmcs = \Pnls-\Cks^2\Plins$.  The resulting smooth fits may have small residuals on small and large wavenumbers due to the broadband difference between $\Psm$ and the original $P_m$. However, we are not correcting for this, as we are only interested in a qualitative rather than quantitative understanding of the shift and any defect in our models of smooth components will reveal itself as a broadband deviation of the non-smooth components from zero. Note that by the definition of $\Pmc$, non-smooth components or a defect in the fit to $\Cks$ will be propagated to $\Pmco$ and $\frac{\Pnlb}{\Plins}- \Cks^2\frac{\Plinb}{\Plins}$. All these should be considered as caveats when we interpret the resulting, non-smooth features.

The black lines in Figure \ref{fig:fPmc} show the source of the observed shift of the acoustic scale, i.e., $\frac{\Pnlb}{\Plins}- \Cks^2\frac{\Plinb}{\Plins}$ in real space. The first three panels show G576 in real space before reconstruction and the middle-right panel shows G576 in real space after reconstruction. The bottom two panels show G576 in redshift space before and after reconstruction. Note that the oscillatory, non-smooth feature in the black lines increases with decreasing redshift, which is consistent with the qualitative trend in Figure \ref{fig:falphas}. The green lines show the non-smooth component in $G(k)$, i.e., $2\Cks^2(\Cko/ \Cks)$, and the blue lines show the non-smooth component in $\Pmc$, i.e., $\Pmco/\Plins$. The non-smooth components of $G(k)$ seem much smaller than that of $\Pmc$, which is consistent with \citet{Pad09}. With $G(k)$ being much smoother than other components, $\Pmco$ appears in good agreement with features in $\frac{\Pnlb}{\Plins}- \Cks^2\frac{\Plinb}{\Plins}$ and therefore indeed seems responsible for the observed shift.

 We do not see a strong trough in $\Pmco$ near $k=0.057\ihMpc$  that corresponds to P-node 1 in Figure 6 of \citet{Crocce08}. This is probably because the actual contribution from this trough is very small as the $\Pmcs/\Plins$ we derived is less than 1/20 at this wavenumber. Considering the errors on the power spectrum (gray error bars) relative to the oscillatory feature and the relatively large contribution to the BAO information from $k=0.1-0.2\ihMpc$ (see \citet{SE03} and also see Figure \ref{fig:StoN} in  \S~\ref{sec:StoN}), we estimate that the off-phase oscillatory feature in $\Pmc$ near $k=0.15\ihMpc$ substantially contributes to the shift. In fact, the mode-coupling shift measured by \citet{Crocce08} for this node (i.e., P-node 4 in their Figure 7) is in good agreement with our measurements of shifts. Features beyond $k \sim 0.2\ihMpc$ are on phase with the BAO, so we do not expect them to contribute to the observed shift. The feature near $k=0.1\ihMpc$ might also be responsible for the shift, but given the deviation of $\Cko$ from zero, it is hard to judge whether the feature is real or due to an imperfect fit to the smooth component.  The middle-right and bottom-right panels show the effect of reconstruction at $z=0.3$ in real space and redshift space, respectively. Oscillatory features in the black and the blue lines have substantially decreased and we no longer find any evident off-phase feature relative to the BAO. The broad-band deviation of components from zero on small wavenumbers in the reconstructed cases is due to a poor fitting to the smooth components.

Figure \ref{fig:fPmc1024} shows the result for G1024 at $z=1$ before reconstruction, in comparison to G576. The different small-scale clustering shown in Figure \ref{fig:fP_Psm}, after accounting for the propagators of G1024 and G576 from \S~\ref{sec:spropagators}, suggests that the mode-coupling term is also slightly different between G1024 and G576. An important question to ask is whether or not there is a difference in the non-smooth component of $\Pmc$ between the two sets, as this will be an indirect way to test the shift difference between G576 and G1024. From Figure \ref{fig:fPmc1024}, there seems to be a resemblance between G576 and G1024 in the oscillatory feature of $\Pmc$ over $k = 0.1-0.2\ihMpc$. Meanwhile the feature is much less significant in the case of G1024 relative to its large sample variance (i.e., the gray error bars) which is consistent with our null detection of shift. We therefore do not find an indication for a difference in the non-smooth component of $\Pmc$ induced by a difference in force and mass resolution or size of the box used for the simulations.

\section{Signal to noise ratio from $G(k)$ and $\Pmc$ }\label{sec:StoN}
\subsection{A toy model}

\begin{figure}[hb]
\vspace{0.4in}
\includegraphics[angle=-90,width=7cm,totalheight=4cm]{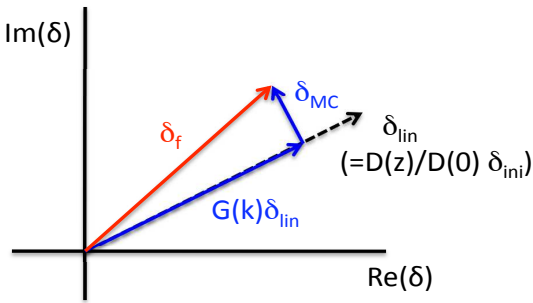}
\caption{A schematic diagram of our toy model to show the Fourier coefficient of nonlinear density field, $\hdelta_f$, relative to the Fourier coefficient of initial, linear density field, $\hdelta_{\rm lin}$ for a given realization. The projection of $\hdelta_f$ onto $\hdelta_{\rm lin}$ is $G(k)\hdelta_{\rm lin}$ and the component uncorrelated to $\hdelta_{\rm lin}$ is $\hdelta_{\rm MC}$.}\label{fig:delta}
\end{figure}

 As pointed out in \citet{PadLag09}, the reconstructed field is not the linear density field at second order. In an attempt to visualize the difference between the linear field and the reconstructed field, let us introduce a toy model in which the density field after reconstruction or before reconstruction is related to the initial density field in the complex plane of Fourier space as following: 
\begin{eqnarray}
\hdelta_f(\vec{k})&=&G(k)\hdelta_{\rm lin}(\vec{k})+\hdelta_{\rm MC}(\vec{k}).
\end{eqnarray}
That is, we describe the final field as a combination of a component along the direction of $\hdelta_{\rm lin}$, $G(k)\hdelta_{\rm lin}$,  and a component that is statistically uncorrelated to the initial field, $\hdelta_{\rm MC}$ (Figure \ref{fig:delta}). Here $G(k)$ is a real-valued propagator and $\hdelta_{\rm lin}$ is a linear density field scaled with a linear growth factor, as before. Note that this toy model is not strictly correct: for example, we are assuming that $\hdelta_{\rm lin}$ and $\hdelta_{\rm MC}$ are statistically independent. Nevertheless, the toy model above returns a consistent definition of a propagator when multiplied with  $\hdelta_{\rm lin}$ and averaged over ensembles:
\begin{eqnarray}
<\hdelta_f(\kvec) \hdelta^*_{\rm lin}(\kvec)>&=&(2\pi)^3G(k)\Plin.
\end{eqnarray}
In other words, the average of the projection of $\hdelta_f$ onto $\hdelta_{\rm lin}$ is $G(k)\Plin$. That is, $G(k)\sim 1$ means that the average projection returns the linear density field.

\begin{figure*}[ht]
\plotone{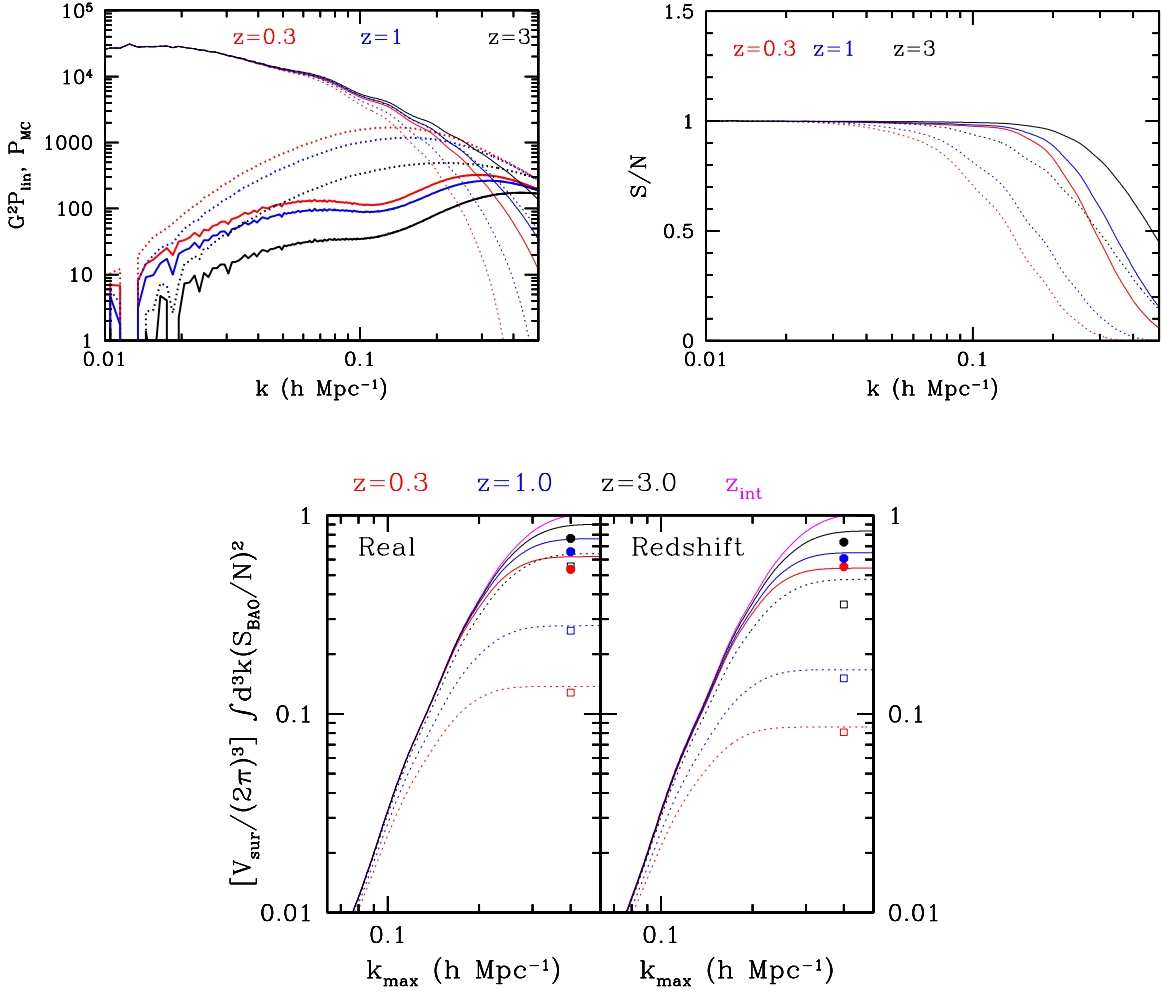}
\caption{Top left: $G(k)^2\Plin$ and $\Pmc$ before (thin and thick dotted lines, respectively) and after reconstruction (thin and thick solid lines, respectively) in redshift space at various redshift. After reconstruction, $G(k)$ increases toward unity and $\Pmc$ decreases. Top right: the corresponding signal-to-noise ratios per Fourier mode in redshift space derived from $G(k)^2\Plin$ and $\Pmc$. Bottom: the total BAO signal-to-noise ratio as a function of the maximum wavenumber $\kmax$. See the text. The data points are ratios of the variance at $z=50$ to the variances at various redshifts from Table \ref{tab:recreal}. }\label{fig:StoN}
\end{figure*}

In this picture, $\Pmc$ represents the random dispersion around the point $G\hdelta_{\rm lin}$:
\begin{eqnarray}
<(\hdelta_f - G\hdelta_{\rm lin})(\hdelta_f-G\hdelta_{\rm lin})^*>&=&<\hdelta_{\rm MC}\hdelta^*_{\rm MC}>,\\
\rightarrow P_f- G^2\Plin & =& \Pmc.
\end{eqnarray}

Therefore, the nonlinear density field, whether before or after reconstruction, can be described by $G(k)$, representing the average projection onto the linear field, and $\Pmc$, representing a random dispersion around the linear field, for a single Fourier component. Obviously, in order to recover the linear density field, we require at least $G(k)=1$ and $\Pmc=0$. In the case of BAO, the recovery of BAO signal is well-described by $G(k)$ and therefore is related to the average projection on the linear field. Meanwhile, it is important to note that the recovery of the broadband shape of the linear power spectrum will be hard to diagnose with $G(k)$ alone, in the presence of $\Pmc$ that also contributes to the broadband shape by definition, and therefore needs to take a different strategy than ones that are effective for the recovery of BAO.

We can relate $G(k)$ and $\Pmc$ to the signal-to-noise ratio of the power spectrum. Since we look for a linear feature in the final power spectrum, the signal, $P_{\rm signal}$, is 
\begin{eqnarray}
<(\hdelta_f - \hdelta_{\rm MC})(\hdelta_f-\hdelta_{\rm MC})^*>&=&G^2(k)<\hdelta_{\rm lin} \hdelta_{\rm lin}^*>\\
\rightarrow P_{\rm signal}(k)&=&G^2(k) \Plin(k),
\end{eqnarray}
as expected and $\Pmc$ contributes to the noise through $P_f$.

\subsection{Signal to Noise ratio}
When assuming a Gaussian error, the signal-to-noise ratio of the standard ruler test, relative to the linear density field, now can be written as
\begin{eqnarray}
\frac{S}{N}&=&\frac{G^2\Plin}{G^2\Plin + \Pmc}=\frac{1}{1+\Pmc/(G^2\Plin)},
\end{eqnarray}
which is valid whether or not we believe the toy model in the previous section.
It is evident that the optimal reconstruction will increase $G(k)$ to unity and decrease the dispersion, $\Pmc$, over the scale of interest.

In the top left panel of Figure \ref{fig:StoN}, we show $G(k)^2\Plin$ and $\Pmc$ before (dotted lines) and after reconstruction (solid lines) in {\it redshift space}, derived from G576. One sees that the reconstruction not only moves the averages of the Fourier modes near the corresponding linear positions, but also decreases the dispersion around the averages. In the top right panel, we show the increase in the signal-to-noise ratio per Fourier mode in redshift space due to reconstruction. For the linear power spectrum, the ratio will be unity over all scales, implying that the signal here does not only incorporate the BAO but all information in $\Plin$. 

We then derive a cumulative quadratic sum of the signal-to-noise ratio after weighing with the available number of Fourier modes at each wavenumber. Also, in order to single out the BAO information in the signal-to-noise ratio, we weigh the signal in $\Plin$ at each wavenumber with the Silk damping effect, i.e.,  $\exp{[-(k R_{\rm Silk})^{1.4}]}/\Plin$ where $R_{\rm Silk}$ is the Silk damping scale, before the quadratic summation \citep[see][for a similar operation]{SE07}: i.e.,
\begin{eqnarray}
\frac{S_{\rm BAO}}{N}&=&\frac{G^2e^{[-(k R_{\rm Silk})^{1.4}]}}{G^2\Plin + \Pmc}.
\end{eqnarray}
In the bottom panel of Figure \ref{fig:StoN}, we show the total BAO signal-to-noise ratio as a function of $\kmax$. The ratio is normalized to unity for the linear power spectrum at $\kmax=0.4\ihMpc$, for convenience. The dependence on $k$ for small wavenumbers is steeper than $k^3$ which is expected from counting the number of available Fourier modes, and this is because  $\exp{[-(k R_{\rm Silk})^{1.4}]}/\Plin$ also increases with $k$ before the exponential damping dominates. Finally we compare these ratios with the variance of the shift at $z=50$ relative to the variances at various redshifts (from Table \ref{tab:recreal}) that are measured in \S~\ref{sec:salphas} using $\kmax=0.4\ihMpc$ (data points). In redshift space, since a spherically averaged P(k) from N-body simulations suffers a larger noise by 7\%, 9\%, and 10\% at z=0.3, 1.0, and 3.0  than when using the full anisotropic signal \citep{Taka09}, we correct for the measured $\sig_\alpha$'s by this factor before comparison (also see \S~\ref{sec:sRyuichi} for more explanation). After the correction, we find less than 12\% discrepancy in errors on shift between the derived signal-to-noise ratio and the measured errors on shifts. 

The role of $\Pmc$ implies that the signal-to-noise ratio will be different for the reconstructed redshift-space power spectrum, depending on whether or not we perform FoG correction (i.e., FoG compression) during reconstruction. As a reminder, the reconstruction tends to stretch FoG further \citep{ESSS07}, as shown in Figure \ref{fig:fP_Psm}. A simple way to reduce this effect is, to compress the spacial dimension along the line of sight by a typical peculiar velocity dispersion of halos relative to the Hubble expansion, identify clusters using an anisotropic friends-of-friends algorithm, and move all the particles to their center of mass of the cluster along the line of sight, decompress the spacial dimension along the line of sight, before conducting reconstruction. We find that, for a moderate level of FoG compression, which recovers the level of FoG effect before reconstruction, $G(k)$ improves slightly, while the increase in $\Pmc$ is more noticeable: as a result, the signal-to-noise ratio worsens slightly (i.e., $\sim 5\%$ effect at $k\sim 0.2\ihMpc$). For more intensive FoG compression to remove most of the FoG effect, the signal-to-noise ratio decreases as much as $\sim 15\%$ at $k\sim 0.2\ihMpc$ despite the improvement in $G(k)$. Therefore, we conclude that, in terms of the signal-to-noise ratio, reconstruction without FoG correction is most effective, although it probably requires more care in the process of an anisotropic fitting due to the strong FoG effect. On the other hand, including a moderate level of FoG correction in the reconstructed power spectrum might improve anisotropic fitting at the expense of total signal-to-noise ratio.

\begin{figure*}[t]
\plotone{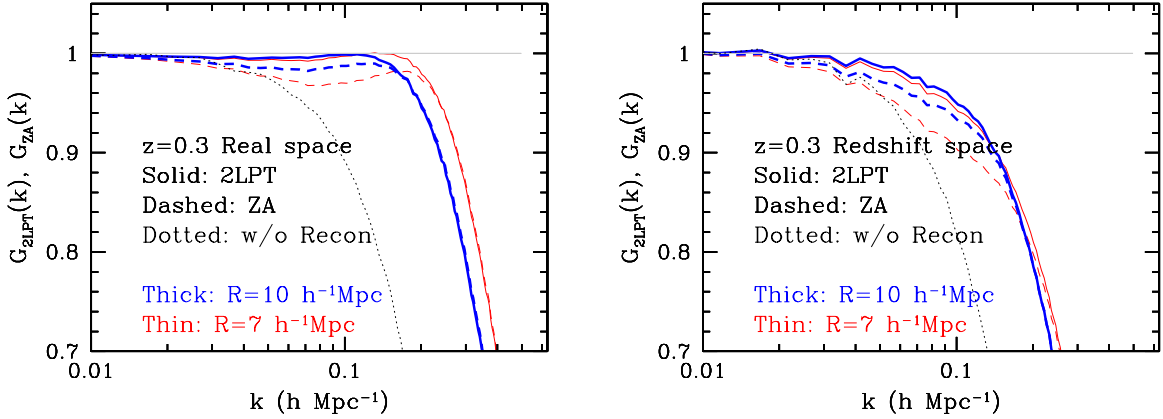}
\caption{Effects of adopting 2LPT into reconstruction (solid lines), relative to our nominal \Zel\ approximation (ZA) based reconstruction (dashed lines). The left panel shows the difference due to 2LPT in real space at $z=0.3$ when we use a Gaussian filter size of $R=10\hMpc$ and $R=7\hMpc$. The right panel shows the effect in redshift space. We use only one $8\trihGpc$ box but rebin $G(k)$ in $\Delta k = 0.005\ihMpc$ in order to decrease the noise in $G(k)$. The dotted line is for $G(k)$ before reconstruction and light gray straight line shows $G(k)=1$. 2LPT seems to correct for the small remaining deviation in $G(k)$ from unity on large scales, which is not likely to make a noticeable change in a signal-to-noise ratio. Note that as we decrease $R$, that is, as we use more nonlinear scale information in the density field to reconstruct BAO, $G(k)$ overall becomes closer to unity on small scales at the expense of more deviations on large scales: 2LPT seems to help fix this deviation on large scales.  }
\label{fig:fCktLPT}
\end{figure*}
\begin{figure*}
\plotone{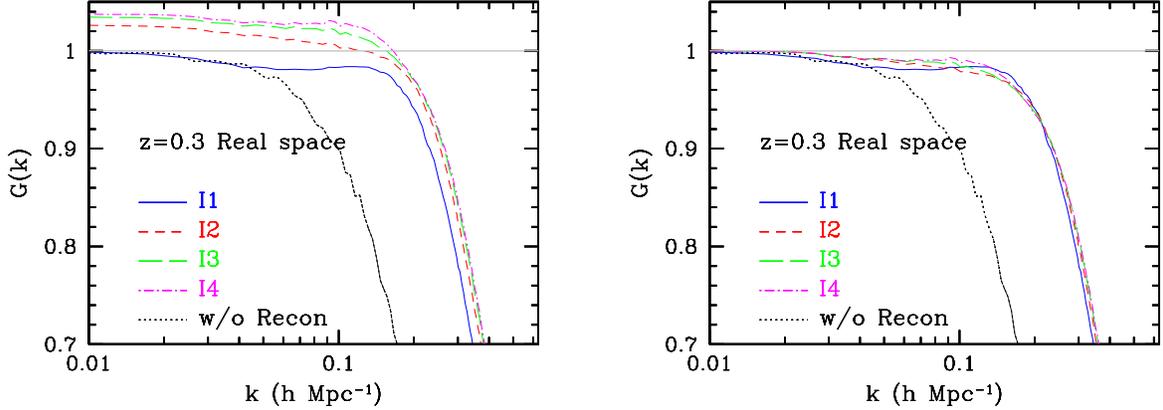}
\caption{Effects of iterative reconstruction. $G(k)$ at $z=0.3$ for various number of iterations. ``I1'' corresponds to our fiducial reconstruction. A filter size of $R=10\hMpc$ is used. We use only one $8\trihGpc$ box but rebin $G(k)$ in $\Delta k = 0.005\ihMpc$ in order to decrease the noise in $G(k)$. The light gray line is for $G(k)=1$, and the dotted black line is $G(k)$ before reconstruction, the colored solid lines are for $G(k)$ for a different number of iterations. In the left panel, $G(k)$ for I2, I3, and I4 is above 1, and it is due to the amplitude of reconstructed power spectrum being slightly larger than the linear theory. The right panel shows the differences after the different amplitude is renormalized. }
\label{fig:fCkIter}
\end{figure*}

\section{Improving reconstruction: 2LPT and Iteration}\label{sec:stLPT}

In the previous sections, we used the reconstruction method presented in \citet{ESSS07} that is based on a linear PT continuity equation. The final density field we use in the continuity equation to derive displacements is, however, nonlinear on small scales, and we therefore smoothed gravity on small scales by using a Gaussian filter. Although most of the displacements are due to large-scale flows \citep{ESSS07}, which is the reason why our fiducial method works well, there must be a small amount of nonlinear contributions to the total displacements: smoothing the small-scale gravity to mimic a linear density field makes one side of the continuity equation linear while, optimally, we want to derive displacements on the other side both due to linear and nonlinear contributions.  In this section, we attempt to improve the performance of reconstruction by adopting a few more operations into our fiducial reconstruction method. We try 2LPT and iteration.

First, we modify the continuity equation based on the second-order Lagrangian perturbation theory. That is, we derive the displacement fields from the following equations \citep[][and references therein]{Sco98}:
\begin{eqnarray}\label{eq:tLPT}
\vec{q}&=&-\bigtriangledown \phi^{(1)} - \frac{3}{7}\bigtriangledown \phi^{(2)},
\end{eqnarray}
where $\vec{q}$ is the estimated displacement fields of the particles from the initial positions, and therefore the reverse of the displacements we will apply to the real particles and reference particles at low redshifts, $\phi^{(1)}$ is the potential in the linear continuity equation derived from linear density field. In our case, we assume that the Gaussian filtered nonlinear density field closely represents the linear density field and derive $\phi^{(1)}$. Then
\begin{eqnarray}
\bigtriangledown^2\phi^{(1)}=\delta_{\rm obs,filtered} 
\end{eqnarray}
and $\phi^{(2)}$ is the second-order correction:
\begin{eqnarray}
\bigtriangledown^2\phi^{(2)}= \sum_{i>j}{\phi_{,ii}^{(1)}\phi_{,jj}^{(1)}-[\phi_{,ij}^{(1)}]^2}.
\end{eqnarray}
The gradient and therefore $\vec{q}$ are evaluated in the comoving (Eulerian) coordinates in our method, and therefore to be exact we are deriving an Eulerian displacement field \citep{Bouchet95}.
In redshift space, we again use equation (\ref{eq:tLPT}), but this time $\phi^{(1)}$ and $\phi^{(2)}$ are derived based on the redshift-space density fields.
As a minor detail, the Gaussian smoothing filter and window function correction due to CIC are applied only once for $\phi^{(2)}$.

As the nonlinear shifts are already close to zero with our fiducial reconstruction method, we focus our attention on the improvement in the signal to noise ratio, that is, the reduction of damping in $G(k)$.  Figure \ref{fig:fCktLPT} shows the effect of adopting 2LPT into our reconstruction for different smoothing lengths, in comparison to our fiducial \Zel\ approximation-based method. As $G(k)$ is the ratio of power spectra from the same random seed, a good portion of the random fluctuations is cancelled in calculating $G(k)$.  We therefore use only one box of G576 simulations. We further reduce the noise by re-binning all power spectra using $\Delta k =0.005\ihMpc$ before generating $G(k)$. 

Overall, the gain of adopting 2LPT is not significant in this specific method: while it pushes $G(k)$ closer to unity on large and quasilinear scales, it is only $\sim 2\%$ effect on $G(k)$ when the filter size $R=10\hMpc$. The figure also shows that a smaller smoothing filter helps reconstruction on small scales while less effective on large scales, probably due to using more nonlinear regions of the density fields. Adopting 2LPT seems to fix the problem. We will further investigate variations in the 2LPT implementation in future work.

Second, we adopt iterative operations into our fiducial reconstruction. At each step in detail, we derive displacement fields based on the density field of real particles ($\delta_{\rm obs}$) at that step and displace the real particles, while the new density field is derived from these displaced particles. At the end of the iterations, we displace the uniformly distributed reference particles by the sum of the displacements from all the iterative steps. Finally we subtract the reference density field ($\delta_{\rm ref, N}$) from the density field of the real particles ($\delta_{\rm obs,N}$) and construct the reconstructed density field ($\delta_{\rm rec}$). The process is schematically shown below:
\begin{eqnarray}
&&\mbox{Step 1:  }\delta_{\rm obs,0} \stackrel{\vec{q}_1=-\bigtriangledown \phi_0}{\longrightarrow}\delta_{\rm obs,1}  \stackrel{\vec{q}_2=-\bigtriangledown \phi_1}{\longrightarrow} \delta_{\rm obs,2} \ldots \nonumber \\
&&\stackrel{\vec{q}_N=-\bigtriangledown \phi_{N-1}}{\longrightarrow} \delta_{\rm obs,N} \nonumber \\
&&\mbox{Step 2:  }\delta_{\rm ref,0} \stackrel{\vec{q}_1+...\vec{q}_N}{\longrightarrow} \delta_{\rm ref,N}\nonumber \\
&&\mbox{Step 3:  }\delta_{\rm rec}=\delta_{\rm obs,N}-\delta_{\rm ref,N} \nonumber
\end{eqnarray}

Figure \ref{fig:fCkIter} shows $G(k)$ as a function of the number of iterative steps: eg., ``I1'' means our fiducial method with $N=1$ in the diagram above. Note that the iterative operations raise the power of the density fields beyond the linear growth, as evident from the constant deviation of $G(k)$ from unity on large scales (left panel). When we renormalize $G(k)$ to match unity on large scales (right panel), the figure implies that iterative reconstructions, at least this specific choice, helps on large scales slightly, then worsens a little bit on the intermediate scales, and then helps on small scales by $\sim 3\%$. It may help to combine 2LPT and the first two steps of iterations (i.e., I2), but, considering the small gain in $G(k)$, we conclude that the extra effort will not be worthwhile.

\section{Testing the Fisher matrix errors}\label{sec:sRyuichi}
Fisher matrix estimates of the uncertainty in the acoustic scale have been useful for designing future BAO surveys, in part due to the simplicity relative to the N-body simulations. Obviously, it is important to ensure, by calibrating the resulting estimates against the \Nb\ study, that the analytic Fisher matrix calculations include reasonable prescriptions for the various nonlinear effects and that it therefore agrees with the real Universe.
However, a precise comparison between the Fisher matrix estimates and the \Nb\ results has been limited by the sample variance, that is, the available number of subsamples. SSEW08 reports a discrepancy of less than $25\%$ between the Fisher matrix estimates based on \citet{SE07} and the \Nb\ results, . The maximum discrepancy is somewhat larger than the expected $1-\sig$ fluctuation on errors of $11\%$ based on their 40 subsamples. Also, the errors from the \Nb\ results tend to be smaller than the Fisher matrix estimates, which is opposite to our intuition that the N-body errors will be larger as there will be nonlinear contribution to the covariance of power spectrum which is not included in the Fisher matrix formalism in \citet{SE07}. This result might be due to estimating the scatter from only 40 simulations. However it was also suspected that there could be a discrepancy of $\gtrsim 1\hMpc$ between the \Zel\ approximation for the nonlinear parameter $\Signl$ and the actual damping in $G(k)$ in their \Nb\ results. However, from Figure \ref{fig:fCk}, the consistency between the two appears better than $\sim 0.5\ihMpc$, except for z=3, which is not enough to account for the discrepancy we observed in SSEW08.

In this section, we again compare the Fisher matrix estimates with the \Nb\ results, T256.  As in SSEW08, we measure the average and the scatters among jackknife subsamples but with 5000 subsamples ($4999\trihGpc$ per subsample) this time. This will allow us to calibrate the formalism in \citet{SE07} with a sample variance of $1/\sqrt{(2\times 5000)}\sim 1\%$, in principle. We find small discrepancies among errors from different resampling methods, and they are in general within their expected uncertainties.

We use WMAP3 cosmology in deriving the Fisher matrix estimates and $\Signl$ from the \Zel\ approximation for the pairs separated by the sound horizon scale. While SSEW08 derive $P_{\rm shot}$ from a numerical shot noise expected from the number density of the tracers, we make a more conservative choice of $P_{\rm shot}$. We derive the constant shot noise contribution  $P_{\rm shot}$ from $\Pnl-D^2(z)\Plin$ at $k=0.2\ihMpc$, where $D(z)$ is the linear growth factor: that is, the derived $P_{\rm shot}$ accounts for both the numerical shot noise and the effects of nonlinear growth near $k=0.2\ihMpc$. Note that our choice of $P_{\rm shot}$ will increase the Fisher matrix error estimates relative to the choice in SSEW08, therefore taking a more conservative stand in resolving the discrepancy reported in  SSEW08. We derive an effective shot noise $n_{\rm eff}=1/P_{\rm shot}$ and list $\nPt$ in  Table \ref{tab:Fisher}.

From Table \ref{tab:Fisher}, the discrepancies in error estimates are within $10\%$ in real space but at the level of $7\%$, $15\%$, $17\%$ in redshift space at  $z=0$, 1, 3, respectively. While we do not observe a consistent trend in real space, the values based on the method in \citet{SE07} tend to be smaller than the measured redshift-space errors. However the measured redshift-space errors may be slightly overestimated for the following reason. In the Fisher matrix method, we assume a fitting to an anisotropic power spectrum, derive errors on $\DA$ and $\hz$, and then project these two dimensional errors on the monopole mode. Meanwhile, in the \Nb\ results, we fit to a spherically averaged redshift-space power spectrum. \citet{Taka09} pointed out that we should expect $\sim 1.2$ times more variance in the monopole power in the latter case  at $z=3$ (i.e., $\sig^2_{P_s} = \sim 1.2 P^2_s/(N_k/2)$ not $P^2_s/(N_k/2)$, see their eq. [9]). That is, fitting to a spherically averaged  power spectrum in redshift space, as we did with the \Nb\ result, does not perform an optimal extraction of a two-dimensional information. We therefore should expect that the scatter from the \Nb\ results in redshift space is about $\sqrt{1.05}=1.02$, $\sqrt{1.16}=1.08$, and $\sqrt{1.2}=1.1$ times more than the Fisher matrix estimates at $z=0$, 1, and 3, respectively. After accounting for this, we end up with less than $10\%$ discrepancies in redshift space as well.
This result is consistent with what we find from the comparison between the cumulative BAO signal-to-noise and the measured errors on shifts, presented in \S~\ref{sec:StoN}, which is reasonable as the Fisher matrix estimates are based on the propagation of errors from power spectrum to the acoustic scale.

 Note that, in real observations, we aim to perform a full two-dimensional analysis and therefore it is legitimate to use the Fisher matrix estimates that come without this extra factor. 
We could further improve the agreements by using a $\Signl$ measured directly from the \Nb\ results, rather than using the analytic \Zel\ approximation, as observed in Figure \ref{fig:fCk}. However, the purpose of Fisher matrix calculations is to bypass the high cost \Nb\ operations, and the \Zel\ approximation has its virtue for being simple and analytic. We therefore conclude that the Fisher matrix formalism in SE07 and the \Nb\ results are in a good agreement and the discrepancy is less than $10\%$ due to an approximation for $\Signl$ and for the covariance matrix in the former.

\section{Conclusion}\label{sec:sdisc}
We summarize the results presented in this paper.

1. We have measured shifts of the acoustic scale due to nonlinear growth and redshift distortions using three sets of simulations, G576, G1024, and T256. which differ in their force, volume, and mass resolution. The measured shifts from the various simulations are in agreement within $\sim 1.5\sigma$ of sample variance. We numerically find $\al(z) -1 = (0.295\pm 0.075)\% [D(z)/D(0)]^{1.74\pm 0.35}$ based on G576. If we fix the power index to be 2, as expected from the perturbation theory, we find the best fit of $\al(z) -1 = (0.300\pm 0.015)\% [D(z)/D(0)]^{2}$. 

2. We find a strong correlation with a non-unity slope between shifts in real space and redshift space. Meanwhile the correlation with the shifts at the initial redshift is weak. Reconstruction not only removes the mean shift and reduces errors on the mean, but also tightens these correlations. After reconstruction, we recover a slope of near unity for the correlation between the shifts in real and redshift space, and restore a strong correlation between the shifts at the low and the initial redshifts. We believe that, as the reconstruction removes shifts due to the second-order, nonlinear process in structure growth, the remaining shifts are dominated by the initial conditions. 
  
3. We find that a propagator is well described by the \Zel\ approximation: for $z \lesssim 1$, we find that the discrepancy, $\Delta \Signl$ is less than $0.5\hMpc$. At high redshift, \Zel\ approximation seems to slightly underestimate the amount of nonlinear damping.

4. We have compared our measurements of shifts from $\chi^2$ fitting with an oscillatory feature in $\Pmc$ and find a qualitative agreement.

5. We construct the signal-to-noise ratio of the standard ruler test based on the measured propagator and mode-coupling term assuming a Gaussian error, and find a consistency with the measured errors on shift. We point out that the propagator describes the average projection of the nonlinear density field onto the linear density field while the mode-coupling term describes any random dispersion uncorrelated to the information of the linear density field. In the case of BAO, the recovery of BAO signal is well-described by $G(k)$, therefore, the average projection on the linear field, while the mode-coupling term contributes to the noise. On the other hand, the recovery of the broadband shape of the linear power spectrum will be hard to characterize with $G(k)$ alone, in the presence of $\Pmc$.

6. We have attempted to improve our reconstruction method by implementing  2LPT and iterative operations. We find only a few \% improvement in $G(k)$. We will further investigate variations in the 2LPT implementation in future work.

7. We test  Fisher matrix estimates of the uncertainty in the acoustic scale using $5000\trihGpc$ of cosmological PM simulations (T256). At an expected sample variance level of 1\%, the agreement between the Fisher matrix estimates based on \citet{SE07} and the \Nb\ results is better than 10 \%.
\\

To conclude this paper, the acoustic peak shifts are small and can be accurately predicted, with control exceeding that required of the observational cosmic
variance limit.  Moreover, reconstruction removes the shifts,
decreases the scatter, and improves the detailed agreement with the
initial density field, as hoped.  We have validated that the acoustic scale shift can be removed to better than 0.02\% for the redshift-space matter field.
We next plan to investigate the effects of galaxy bias (Mehta et al.,
in preparation).
\\

\acknowledgements
We thank Martin Crocce for useful conversations.
H-JS is supported by the U.S. Department of Energy under contract No. DE-AC02-07CH11359.
JE, DJE, KM, and XX were supported by NSF AST-0707725 and 
by NASA BEFS NNX07AH11G.  RT is supported by Grant-in-Aid for Scientific 
  Research on Priority Areas No.467 ``Probing the Dark Energy 
  through an Extremely Wide and Deep Survey with Subaru Telescope''.
MW was partially supported by 
NASA BEFS NNX07AH11G. 




\clearpage

\newcommand{\tableskip}{\\[-8pt]}
\newcommand{\singleline}{\tableskip\hline\tableskip}
\newcommand{\doubleline}{\tableskip\hline\tableskip}
\newlength{\tablespread}\setlength{\tablespread}{30pt}
\newcommand{\dje}{\hspace{\tablespread}}
\tabletypesize{\small}
\def\arraystretch{1.1}

\begin{center}
\begin{deluxetable}{cccccc}
\tablewidth{0pt}
\tabletypesize{\footnotesize}
\tablecaption{\label{tab:tabsim} \Nb\ sets used in this paper.}
\startdata \hline\hline
Sample& Box size & $N_{\rm particles}$ & Force resolution& Total Volume  & Code \\
&($\hGpc$)& & ($\hMpc$) & ($\trihGpc$) & \\ \hline
G576  & 2	 & $576^3$  & 0.1736 & 504 & Metchnik \& Pinto\\
G1024 & 1        & $1024^3$ & 0.0488 & 44  & Metchnik \& Pinto\\
T256  & 1        & $256^3$  & 3.9062 & 5000 & Gadget-2 PM 
\enddata
\tablecomments{Parameters of the simulations used in this paper.}
\end{deluxetable}
\end{center}

\begin{center}
\begin{deluxetable}{cccccc}
\tablewidth{0pt}
\tabletypesize{\footnotesize}
\tablecaption{\label{tab:recreal} The evolution of $\al$.}
\startdata \hline\hline
\multicolumn{6}{c}{G576 before Reconstruction} \\ \hline
    &$\al_{\rm r}-1 (\%)$   &$\al_{\rm s}-1 (\%)$    & $\Delta \al_{sr} (\%)$ & $\Delta \al_{\rm r,z50} (\%)$ &$\Delta \al_{\rm s,z50} (\%)$ \\\hline
0.3&$0.2283\pm 0.0609$      &$0.2661 \pm 0.0820$   &$0.03774\pm 0.0375$ &$0.2272\pm 0.0550$ &$0.2650\pm 0.0781$\\
1.0&$0.1286\pm 0.0425$      &$0.1585 \pm 0.0611$   &$0.0299\pm 0.0323$ &$0.1275\pm 0.0341$ &$0.1574 \pm 0.0559$\\
3.0&$0.0435\pm 0.0293$      &$0.0582 \pm 0.0402$   &$0.0147\pm 0.0213$ &$0.0424\pm 0.0175$ &$0.0571 \pm 0.0318$\\
50.0&$0.001\pm 0.0218$      &&&&\\\hline
\multicolumn{6}{c}{G576 after Reconstruction} \\ \hline
0.3&$-0.0037\pm 0.0298$     &$-0.0015\pm 0.0314$   &$0.0022\pm 0.0128$ &$-0.0048\pm 0.0155$ &$-0.0026\pm 0.0194$ \\
1.0&$-0.0037\pm 0.0269$     &$-0.0024\pm 0.0305$   &$0.0013\pm 0.0128$ &$-0.0048\pm 0.0113$ &$-0.0035\pm 0.0182$\\
3.0&$-0.0021\pm 0.0249$     &$-0.0058\pm 0.0280$   &$-0.0037\pm 0.0125$ &$-0.0032\pm 0.0074$ &$-0.0069\pm 0.0145$ \\\hline\hline
\multicolumn{6}{c}{G1024 before Reconstruction} \\ \hline
1.0&$-0.112\pm 0.163$      &$0.002\pm  0.233$   &$0.114\pm 0.113$ &$-0.142 \pm 0.144$ &$-0.028\pm 0.226$\\
50.0&$0.030\pm 0.088$ & & & & \\ \hline
\multicolumn{6}{c}{G1024 after Reconstruction} \\ \hline
1.0&$-0.093\pm 0.094$      &$-0.055\pm 0.101$     &$0.039\pm 0.043$  &$-0.123 \pm 0.044$  &$-0.085\pm 0.068$
\enddata
\tablecomments{Fitting range: $0.02\ihMpc \leq k \leq 0.4\ihMpc$. We use $504 \trihGpc$ for G576 and $44\trihGpc$ for G1024. A subscript ``r''means a value in real space and ``s'' means a value in redshift space. }
\end{deluxetable}
\end{center}

\begin{deluxetable}{c|c|ccc|ccc} 
\tablewidth{0pt}
\tabletypesize{\small}
\tablecaption{\label{tab:Fisher} The Fisher matrix estimates in comparison to the N-body data from T256.}
\startdata \hline\hline
\multicolumn{1}{c}{} & \multicolumn{1}{c}{}  & \multicolumn{3}{|c}{\Nb\ data} &\multicolumn{3}{|c}{Fisher matrix} \\ \hline
&$z$ &Sample &$\Signl$ &         $\sig_\al (\%)$    &$\Signn$  &  $\nPt$ & $\sial$ \\\hline
Real space & 0.0& T256 & 8.27    &0.0217   
&8.27  &4.38 & 0.0234\\
&1.0& T256 &5.26            &0.0133
&5.26  &10.6 & 0.0136 \\
&3.0& T256 &2.78            &0.0090
&2.78 & 38.0 & 0.0081\\
&20& T256 &0.0             &0.0069
 &0.53 & 100 &0.0072\\\hline
&$z$  &Sample & $\Signl$   &       $\sig_\al (\%)$      & $\Sigxy/\Sigz$ &  $\nPt(\mu=0)$ & $\sial$ \\ \hline
Redshift space&0.0&T256    &10.0         &0.0274  
& 8.27/11.76 &4.38 & 0.0257\\
&1.0&T256 &7.0        &0.0192
&5.26/9.55  &10.6 & 0.0167\\
&3.0&T256 &4.0       &0.0117
&2.78/5.47    &38.0  & 0.0100 \\\hline
\enddata
\tablecomments{N-body results are derived by using a total 5000 jackknife samples.
Values of $\Signl$ in the fourth column represents the nonlinear smoothing used for the template $P_m(k)$ in the $\chi^2$ analysis of the \Nb\ data, and $\Signn$ and $\Sigxy/\Sigz$ in the sixth column are derived from the \Zel\ approximations and represents the nonlinear degradation of the BAO assumed in the Fisher matrix calculations.}
\end{deluxetable}


\begin{thebibliography}{}\frenchspacing


\bibitem[Angulo et al.(2005)]{Angulo05} Angulo, R., Baugh,
C.~M., Frenk, C.~S., Bower, R.~G., Jenkins, A., \& Morris, S.~L.\ 2005,
\mnras, 362, L25

\bibitem[Angulo et al.(2008)]{Angulo08} Angulo, R.~E., Baugh,
C.~M., Frenk, C.~S., \& Lacey, C.~G.\ 2008, \mnras, 383, 755

\bibitem[Bennett et al.(2003)]{BennettWmap} Bennett, C.~L., et al.\
2003, \apjs, 148, 1


\bibitem[Beno{\^ i}t et al.(2003)]{BenoitArcheops} Beno{\^ i}t, A.~et
al.\ 2003, \aap, 399, L19

\bibitem[de Bernardis et al.(2000)]{deB00}  de Bernardis, P., et al., 2000, Nature, 404, 955



\bibitem[Blake \& Glazebrook(2003)]{Blake03} Blake, C., \&
Glazebrook, K.\ 2003, \apj, 594, 665

S.\ 2005, \mnras, 363, 1329

Bassett, B., Glazebrook, K., Kunz, M., \& Nichol, R.~C.\ 2006, \mnras, 365,
255

\bibitem[Blake et al.(2007)]{Blake07} Blake, C., Collister, A.,
Bridle, S., \& Lahav, O.\ 2007, \mnras, 374, 1527

\bibitem[Bond \& Efstathiou(1984)]{Bond84}
        Bond, J.R. \& Efstathiou, G. 1984,
        \apj, 285, L45

\bibitem[Bouchet et 
al.(1995)]{Bouchet95} Bouchet, F.~R., Colombi, S., Hivon, E., \& Juszkiewicz, R.\ 1995, \aap, 296, 575 


\bibitem[Cole et al.(2005)]{Cole05} Cole, S., et al.\ 2005,
\mnras, 362, 505



\bibitem[Crocce \& Scoccimarro(2006)]{Crocce06b} Crocce, M., \&
Scoccimarro, R.\ 2006, \prd, 73, 063520

\bibitem[Crocce \& Scoccimarro(2008)]{Crocce08} Crocce, M., \& Scoccimarro, R.\ 2008, \prd, 77, 023533



T


\bibitem[Eisenstein \& Hu(1998)]{EH98} Eisenstein, D.~J., \&
Hu, W.\ 1998, \apj, 496, 605


\bibitem[Eisenstein(2003)]{Eisen03}
        Eisenstein, D.J., 2003, in ASP Conference Series, volume 280, Next Generation Wide Field Multi-Object Spectroscopy,
        ed. M.J.I. Brown \& A. Dey (ASP: San Francisco) pp. 35-43;
        astro-ph/0301623


\bibitem[Eisenstein et al.(2005)]{Eisen05} Eisenstein, D.~J.,
et al.\ 2005, \apj, 633, 560

\bibitem[Eisenstein et al.(2007)]{ESSS07} Eisenstein, D.~J.,
Seo, H.-J., Sirko, E., \& Spergel, D.~N.\ 2007, \apj, 664, 675

\bibitem[Eisenstein et al.(2007)]{ESW07} Eisenstein, D.~J.,
Seo, H.-J., \& White, M.\ 2007, \apj, 664, 660


\bibitem[Estrada et al.(2009)]{Estra08} Estrada, J., Sefusatti,
E., \& Frieman, J.~A.\ 2009, \apj, 692, 265


\bibitem[Gazta{\~n}aga et al.(2009)]{Gazt08a} Gazta{\~n}aga,
E., Cabr{\'e}, A., Castander, F., Crocce, M.,
\& Fosalba, P.\ 2009, \mnras, 399, 801
2480 

\bibitem[Gazta{\~n}aga et al.(2009)]{Gazt08b} Gazta{\~n}aga,
E., Miquel, R.,
\& S{\'a}nchez, E.\ 2009, Physical Review Letters, 103, 091302


\bibitem[Halverson et al.(2002)]{HalDasi}
        Halverson, N.~W.~et al.\ 2002, \apj, 568, 38




\bibitem[Hanany et al.(2000)]{Han00}
        Hanany, S., et al., 2000,
        \apj, 545, L5

\bibitem[Hinshaw et al.(2007)]{Hinshaw07} Hinshaw, G., et al.\
2007, \apjs, 170, 288

\bibitem[Hinshaw et al.(2009)]{Hinshaw08} Hinshaw, G., et al.\
2009, \apjs, 180, 225



\bibitem[Holtzman(1989)]{Holtzman89} Holtzman, J.~A.\ 1989, \apjs,
71, 1


\bibitem[Hu \& Sugiyama(1996)]{HS96} Hu, W., \& Sugiyama,
N.\ 1996, \apj, 471, 542

\bibitem[Hu \& Haiman(2003)]{Hu03} Hu, W., \& Haiman, Z.\
2003, \prd, 68, 063004

\bibitem[Hu \& White(1996)]{Hu96} Hu, W., \& White, M.\ 1996, \apj, 471, 30

\bibitem[Huff et al.(2007)]{Huff07} Huff, E., Schulz, A.~E.,
White, M., Schlegel, D.~J.,
\& Warren, M.~S.\ 2007, Astroparticle Physics, 26, 351

\bibitem[H{\"u}tsi(2006)]{Hutsi06} H{\"u}tsi, G.\ 2006, \aap,
449, 891


\bibitem[Jeong \& Komatsu(2006)]{Jeong06} Jeong, D., \&
Komatsu, E.\ 2006, \apj, 651, 619

\bibitem[Jeong \& Komatsu(2009)]{Jeong09} Jeong, D., \& Komatsu, E.\ 2009, \apj, 691, 569 


\bibitem[Kaiser(1987)]{Kaiser87} Kaiser, N.\ 1987, \mnras, 227,
1

\bibitem[Kazin et al.(2010)]{Kazin09} Kazin, E.~A., et al.\
2010, \apj, 710, 1444


\bibitem[Komatsu et al.(2009)]{Komatsu09} Komatsu, E., et al.\ 
2009, \apjs, 180, 330 


\bibitem[Lee et al.(2001)]{Lee01} Lee, A.~T., et al.\ 2001,
\apjl, 561, L1

\bibitem[Linder(2003)]{Linder03} Linder, E.~V.\ 2003, \prd, 68,
083504



\bibitem[Mohayaee et al.(2006)]{MAK06} Mohayaee, R., Mathis, 
H., Colombi, S., \& Silk, J.\ 2006, \mnras, 365, 939 

\bibitem[Matarrese \& Pietroni(2007)]{Matarrese07} Matarrese, S., \& Pietroni, M.\ 2007, JCAP, 6, 26

\bibitem[Matsubara(2008)]{Mat08} Matsubara, T.\ 2008, \prd,
77, 063530

\bibitem[Mehta et al. (2010)]{Mehta09} Mehta et al. in preparation

\bibitem[Meiksin, White, \& Peacock(1999)]{Meiksin99}
        Meiksin, A., White, M., \& Peacock, J.~A.\ 1999, \mnras, 304, 851



\bibitem[Metchnik \& Pinto(2010)]{Metchnik} Metchnik \& Pinto in preparation

\bibitem[Miller et al.(1999)]{Mil99}
        Miller, A.D., Caldwell, R., Devlin, M.J., Dorwart, W.B., Herbig, T.,
        Nolta, M.R., Page, L.A., Puchalla, J., Torbet, E., \& Tran, H.T., 1999,
        \apj, 524, L1


\bibitem[Netterfield et al.(2002)]{Netter02} Netterfield, C.~B.,
et al.\ 2002, \apj, 571, 604

\bibitem[Nishimichi et al.(2007)]{Nishimichi07} Nishimichi, T., et
al.\ 2007, \pasj, 59, 1049

\bibitem[Noh et al.(2009)]{Noh09} Noh, Y., White, M.,
\& Padmanabhan, N.\ 2009, \prd, 80, 123501


\bibitem[Okumura et al.(2008)]{Okumura08} Okumura, T., Matsubara,
T., Eisenstein, D.~J., Kayo, I., Hikage, C., Szalay, A.~S.,
\& Schneider, D.~P.\ 2008, \apj, 676, 889

\bibitem[Padmanabhan et al.(2007)]{Pad07} Padmanabhan, N., et
al.\ 2007, \mnras, 378, 852

\bibitem[Padmanabhan
\& White(2009)]{Pad09} Padmanabhan, N., \& White, M.\ 2009, \prd, 80, 063508


\bibitem[Padmanabhan et al.(2009)]{PadLag09} Padmanabhan, N., 
White, M., \& Cohn, J.~D.\ 2009, \prd, 79, 063523 




\bibitem[Pearson et al.(2003)]{Pearson03} Pearson, T.~J., et al.\
2003, \apj, 591, 556

\bibitem[Peebles \& Yu(1970)]{Peebles70} Peebles, P.~J.~E.~\& Yu,
J.~T.\ 1970, \apj, 162, 815


\bibitem[Percival et al.(2007a)]{Percival07a} Percival, W.~J., et
al.\ 2007, \apj, 657, 51

\bibitem[Percival et al.(2007b)]{Percival07b} Percival, W.~J., Cole,
S., Eisenstein, D.~J., Nichol, R.~C., Peacock, J.~A., Pope, A.~C.,
\& Szalay, A.~S.\ 2007, \mnras, 381, 1053

\bibitem[Percival et al.(2010)]{Percival09} Percival, W.~J., et
al.\ 2010, \mnras, 401, 2148


\bibitem[S{\'a}nchez et al.(2008)]{Sanchez08} S{\'a}nchez, A.~G., 
Baugh, C.~M., \& Angulo, R.\ 2008, \mnras, 390, 1470 

\bibitem[S{\'a}nchez et al.(2009)]{Sanchez09} S{\'a}nchez, A.~G.,
Crocce, M., Cabr{\'e}, A., Baugh, C.~M.,
\& Gazta{\~n}aga, E.\ 2009, \mnras, 400, 1643


\bibitem[Scoccimarro(1998)]{Sco98} Scoccimarro, R.\ 1998, 
\mnras, 299, 1097 


\bibitem[Seo \& Eisenstein(2003)]{SE03} Seo, H.-J., \&
Eisenstein, D.~J.\ 2003, \apj, 598, 720

\bibitem[Seo \& Eisenstein(2005)]{SE05} Seo, H.-J., \&
Eisenstein, D.~J.\ 2005, \apj, 633, 575

\bibitem[Seo \& Eisenstein(2007)]{SE07} Seo, H.-J., \&
Eisenstein, D.~J.\ 2007, \apj, 665, 14

\bibitem[Seo et al.(2008)]{SSEW08} Seo, H.-J., Siegel, E.~R., 
Eisenstein, D.~J., \& White, M.\ 2008, \apj, 686, 13



\bibitem[Sirko(2005)]{Sirko05} Sirko, E.\ 2005, \apj, 634, 728

\bibitem[Smith et al.(2003)]{Smith03} Smith, R.~E., et al.\ 
2003, \mnras, 341, 1311 

\bibitem[Smith et al.(2007)]{Smith07} Smith, R.~E.,
Scoccimarro, R., \& Sheth, R.~K.\ 2007, \prd, 75, 063512

\bibitem[Smith et al.(2008)]{Smith08} Smith, R.~E.,
Scoccimarro, R., \& Sheth, R.~K.\ 2008, \prd, 77, 043525

\bibitem[Spergel et al.(2007)]{Spergel07} Spergel, D.~N., et al.\ 
2007, \apjs, 170, 377 

\bibitem[Springel et al.(2001)]{SpringelGadget} Springel, V., Yoshida, 
N., \& White, S.~D.~M.\ 2001, New Astronomy, 6, 79 


\bibitem[Springel et al.(2005)]{Springel05} Springel, V., et al.\
2005, \nat, 435, 629

\bibitem[Springel(2005)]{SpringelGadget2} Springel, V.\ 2005, \mnras, 
364, 1105 


\bibitem[Sunyaev \& Zeldovich(1970)]{SZ70} Sunyaev, R.~A., \& Zeldovich, Y.~B.\ 1970, \apss, 7, 3

\bibitem[Takahashi et al.(2008)]{Taka08} Takahashi, R., et
al.\ 2008, \mnras, 389, 1675

\bibitem[Takahashi et al.(2009a)]{Taka09} Takahashi, R., et
al.\ 2009, \apj, 700, 479


\bibitem[Takahashi et al.(2009)]{Taka10} Takahashi, R., et 
al.\ 2009, arXiv:0912.1381 

\bibitem[Taruya et al.(2009)]{Taruya09} Taruya, A., Nishimichi,
T., Saito, S., \& Hiramatsu, T.\ 2009, \prd, 80, 123503

\bibitem[Tegmark et al.(2006)]{Tegmark06} Tegmark, M., et al.\
2006, \prd, 74, 123507



\bibitem[White(2005)]{White05} White, M.\ 2005, Astroparticle
Physics, 24, 334

\bibitem[Xu et al.(2010)]{Xu09} Xu, X., et al.\ 2010, \apj,
718, 1224


\bibitem[Zel'dovich(1970)]{Zel70}
        Zel'dovich, Y.A., 1970, \aap, 5, 84



\end{thebibliography}
\end{document}